\pdfoutput=1
\documentclass[acmsmall,nonacm]{acmart}
\usepackage{amsmath, amsthm}
\usepackage{balance}

\usepackage{hyperref}
\usepackage{natbib}
\usepackage{cleveref}

\usepackage{float}
\usepackage{listings}
\usepackage{placeins}
\usepackage{subcaption}
\usepackage{wrapfig}
\usepackage{booktabs}
\usepackage{tabularx}
\usepackage[most]{tcolorbox}

\usepackage{tikz}
\usetikzlibrary{automata, arrows.meta, positioning}
\usepackage{pgf-pie}  
\usepackage{geometry}
\usepackage{graphicx}
\usepackage{pdflscape}
\usepackage{pgfplots}
\usepackage{pgfplotstable}
\usetikzlibrary{positioning}
\pgfplotsset{compat=newest}
\usepgfplotslibrary{units, external}
\usepackage{filecontents}
\usepackage{stmaryrd}

\usepackage{todonotes}
\usepackage{subfiles}
\newtheorem{definition}{Definition}[section]
\newtheorem{example}{Example}[section]

\usepackage{algpseudocode}
\usepackage{algorithm}
\usepackage[acronym]{glossaries}
\usepackage[T1]{fontenc}
\usepackage{semantic}

\newcommand{\runa}[1]{\textsc{(#1)}}
\newcommand{\cat}[1]{\textbf{#1}}

\newcommand{\loc}{\ensuremath{\mathscr{l}}}

\newcommand{\Exp}{\ensuremath{\mathbb{E}xp}}

\newcommand{\Var}{\ensuremath{\mathbb{V}ar}}

\newcommand{\Pow}[1]{\ensuremath{\mathbb{P}(#1)}}

\algnewcommand\algorithmicforeach{\textbf{for each}}
\algdef{S}[FOR]{ForEach}[1]{\algorithmicforeach\ #1\ \algorithmicdo}

\acmJournal{JACM}
\acmVolume{1}
\acmNumber{1}
\acmArticle{1}
\acmMonth{1}

\setcopyright{rightsretained}
\acmPrice{0.00}
\acmDOI{}

\begin{document}
	%\subfile{frontmatter/resume.tex}
   %Frontpage stuff
\title[A type system for data flow analysis]{A type system for data
  flow and alias analysis in ReScript}

\author{Nicky Ask Lund} 
\email{nlund18@student.aau.dk} 
\affiliation{%
	\institution{Aalborg University}
	\department{Department of Computer Science}
	\city{Aalborg}
	\country{Denmark}
}

\author{Hans Hüttel} 
\email{hans.huttel@di.ku.dk} 
\affiliation{%
	\institution{University of Copenhagen}
	\department{Department of Computer Science}
	\city{Copenhagen}
	\country{Denmark}
}

\date{August 2023}


\begin{abstract}
	ReScript is a strongly typed language that targets
        JavaScript, as an alternative to gradually typed languages,
        such as TypeScript. In this paper, we present a sound type system
        for data-flow analysis for a subset of the ReScript language,
        more specifically for a $\lambda$-calculus with mutability and
        pattern matching. The type system is a local analysis that
        collects information about variables are used at each program
        point as well as alias information. 
\end{abstract}

\maketitle
\renewcommand{\shortauthors}{}

%%% Local Variables:
%%% mode: latex
%%% TeX-master: "../pepm2024"
%%% End:

	\tableofcontents

\section{Introduction}
Data-flow analysis has been studied for decades to better to provide flow information of programs.
This flow information has been used for different tasks for compiler optimization, debugging and understanding programs, testing and maintenance.
In the context of compiler optimization, where the flow information provides data that may be used at given parts of the program at runtime.

The classical way of doing data-flow analysis has been by using iterative algorithm based on representing the control-flow of programs as graphs.
The purpose of such graphs is to give a sound over-approximation of the control flow of a program, where edges represent the flow and nodes represent basic blocks.
By using the information of control-flow graphs, many algorithms have been developed to use those by annotating the graphs and solving the maximal fix-point\cite{KildallGaryA1973Auat, RyderBarbara1988Idaa}. (Ref to algorithms, such as kilders)
Other techniques have also been presented, such as a graph-free approach \cite{HorspoolR.Niegel2002AGAt} or through type systems with refinement types \cite{PavlinovicZvonimir2021Dfrt}.

When analysing languages, such as C/C++ or other languages that explicitly handles pointers, it is important to take into account aliasing, i.e., multiple variables referring to the same location.
Many data-flow analysis uses alias algorithms to compute this information.
Two overall types of alias algorithm has been used, flow-sensitive which give precise information but are expensive, and flow-insensitive which are less precise but are inexpensive \cite{LiangDonglin1999Eaag, EmamiMaryam1994Cipa}.
\bigskip

This paper will focus on data-flow analysis with focus on a subset of the functional language ReScript, a new language based on OCaml with a JavaScript inspired syntax which targets JavaScript.
ReScript offers a robust type system based on OCaml, which provides an alternative to other gradually typed languages that targets JavaScript.\cite{rescript_rebrand}

As ReScript provides integration with JavaScript, it provides its own compiler toolchain and build system for optimizing and compile to JavaScript.
ReScript does, however, introduce an analysis tool for dead-code, exception, and termination analysis, but the tool is only experimental.\cite{reanalyze}
As ReScript introduces mutability, through reference constructs for creation, reading, and writing,

We will the present a type system for data-flow analysis for bindings and alias analysis.
As type system have been used to provide a semantic analysis of programs usually used to characterize specific type of run-time errors.
Type systems are implemented as either static or dynamic analysis, i.e., on compile time or run-time.
Type systems are widely used, from weakly typed languages such as JavaScript, to strongly typed languages commonly found in functional languages such as Haskell and Ocaml.

This work is a generalization of \cite{DVNicky}, which focused on dead-value analysis.
We will present the analysis for a language based on a subset of the ReScript language, for a $\lambda$-calculus with mutability, local bindings, and pattern matching.
The type system we proposes provides the data information used at each program point and the alias information used.
Since the analysis we present focus collecting dependencies that are used to evaluate a part of a program, we present a local analysis of programs.
\bigskip

We will first present the language, its syntax and semantics, in \cref{sec:lang} and the type system for data-flow analysis in \cref{sec:TypeSys}.
Then we will present the soundness of the type system in \cref{sec:Soundness}, and lastly we will conclude in \cref{sec:Conc}.

\section{Language}\label{sec:lang}
This section will introduce a functional programming language, based on a subset of ReScript.
As this is a generalization of a dead-value analysis system, the language presented here is based on the one found in \cite{DVNicky}.
The language we present is basically a $\lambda$-calculus with bindings, pattern matching and mutability.
As the purpose of the dependency analysis is to analyse each subexpression of a program and differentiate them, the language is extended with labelling, which we also call program points, all expressions and subexpressions.
When labelling a syntactical element or semantic element, we call it an occurrence, such that the analysis is done for an occurrence and its sub-occurrences, while the semantic occurrences are variables and locations.

In the language we assume that all local bindings, and recursive bindings, are unique, which can be ensured by using $\alpha$-conversion on an occurrence.
We also make a distinction between labelled and unlabelled expressions, such that we call occurrences as labelled expression, and we call unlabelled expressions as expressions.

In this section we will first formally introduce the abstract syntax for the language, where we will then present binding models.
Then we will present the dependency function, to model the semantic flow-data, and lastly we will present the semantic as a big-step operational semantics.

\subsection{Syntax}
This section introduces the abstract syntax of the language, based on the one presented in \cite{DVNicky}.
The syntactic categories for the language is defined as:

\begin{align*}
	p\in &\;\cat{P} &-\;&\mbox{The category for program points} \\
	e\in &\;\cat{Exp} &-\;&\mbox{The category for expressions, or unlabelled occurrencens} \\
	o\in &\;\cat{Occ} &-\;&\mbox{The category for occurrences, or labelled expressions} \\
	c\in &\;\cat{Con} &-\;&\mbox{The category for constants} \\
	x,\;f\in &\;\cat{Var} &-\;&\mbox{The category for variables} \\
	\loc\in &\;\cat{Loc} &-\;&\mbox{The category for constants}
\end{align*}

We also introduce a notation for occurrences of categories where, for a category $cat$, we write $cat^\cat{P}$ to denote the pair $cat\times\cat{P}$, for occurrences as such:
$\cat{Exp}^\cat{P}=\cat{Exp}\times\cat{P}$.

Since the category for occurrences are labelled expressions, it can further be defined as:
$$\cat{Occ}=\cat{Exp}^\cat{P}$$

The formation rules is then presented in \cref{fig:coresyntax}.

\begin{figure}[H]
	\begin{minipage}[t]{0.45\textwidth}
		\setlength\tabcolsep{4pt}
		\begin{tabular}{>{$}l<{$}>{$}r<{$}>{$}l<{$}}
			Occurrence \; o &::= &e^p \\\\

			expression \; e &::= &x \mid c \mid o_1\;o_2 \mid \lambda x.o\\
			&| &c \; o_1 \; o_2\\
			&| &\mbox{let} \; x \; o_1 \; o_2 \\
			&| &\mbox{let rec} \; x \; o_1 \; o_2 \\
			&| &\mbox{case} \; o_1 \; \tilde{\pi} \; \tilde{o}\\
			&| &\mbox{ref} \; o \mid o_1 := o_2 \mid \; !o\\\\

			Pattern \; s &::= &n \mid b \mid x \mid \_
		\end{tabular}
	\end{minipage}
	\begin{minipage}[t]{0.45\textwidth}
		\setlength\tabcolsep{4pt}
		\begin{tabular}{>{$}l<{$}>{$}r<{$}>{$}l<{$}}
			Constant\; c &::= &n \mid b\\
			&| &PLUS \\
			&| &MINUS \\
			&| &TIMES \\
			&| &EQUAL \\
			&| &LESS\\
			&| &GREATER\\ \\

			Patterns		\; \tilde{\pi} &::= &(s_1,\cdots,s_n)\\\\

			Occurrences \; \tilde{o} &::= &(e_1^{p_1},\cdots,e_n^{p_n})
		\end{tabular}
	\end{minipage}
	\caption{Abstract syntax}
	\label{fig:coresyntax}
\end{figure}

Some notable constructs is further explained below.
\begin{description}
	\item[Abstractions] $\lambda\;x.o$ denotes an abstraction, with a parameter $x$ and body $o$.

	\item[Constants] $c$ are either natural numbers $n$, boolean values $b$, or functional constants.
		We introduce a function $apply$, that for each functional constant $c$ returns the result of applying $c$ to its arguments.
		$$apply(PLUS,2,2)=2+2$$

	\item[Bindings] $\mbox{let} \; x \; o_1 \; o_2$ and $\mbox{let rec} \; f \; o_1 \; o_2$, also called local declarations, are immutable bindings that binds the variables $x$ to values $o_1$ evaluates to.
		We also introduces non-recursive and recursive bindings, by using the \emph{rec} keyword.

	\item[Reference] $\mbox{ref\;o}$ is the construct for creating references which are handled as locations and allows for binding locations to local declarations.
		We also introduces constructs for reading from references, $!o$, and writing to references, $o_1\;:=\;o_2$.

	\item[Pattern matching] $\mbox{case} \; o_1 \; \tilde{\pi} \; \tilde{o}$, matches an occurrence with the ordered set, $\tilde{\pi}$, of patterns.
		For each pattern in $\tilde{\pi}$ there is also an occurrence in $\tilde{o}$, as such, both sets must be of equal size.
		We also denote the size of patterns as $|\tilde{\pi}|$ and the size of occurrences as $|\tilde{o}|$.
\end{description}

\begin{example}[]\label{ex:write}
Consider the following occurrence:
\begin{lstlisting}[language=Caml, mathescape=true]
(let x (ref 3$^1$)$^2$ (let y (let z (5$^3$)$^4$ (x$^5$:=z$^7$)$^8$)$^{9}$ (!x)$^{10}$)$^{11}$)$^{12}$
\end{lstlisting}
Here, we first creates a reference to the constant 3 and binds this reference to $x$ (Such that $x$ is an alias of this reference).
Secondly we create a binding for $y$, where create a binding $z$, to the constant 5, before writing to the reference, that $x$ is bound to, to the value that $z$ is bound to.
Lastly, we read the reference that $x$ is bound to, where we expect to retrieve the value $5$.
\end{example}

Next we defined the notion of free variables, in the usual way for $\lambda$-calculus, as follows:
\begin{definition}[Free variables]\label{def:fv}
	The set of free variables is a function $fv:\cat{Occ}\rightarrow\Pow{\cat{Var}}$, given inductively by:
	\begin{align*}
		fv(x^p)&=\{x\}\\
		fv(c^p)&=\emptyset\\
		fv([\lambda\;y.e^{p'}]^p)&=fv(e^{p'})\backslash\{y\}\\
		fv([e_1^{p_1}\;e_2^{p_2}]^p)&=fv(e_1^{p_1})\cup fv(e_2^{p_2})\\
		fv([c\;e_1^{p_1}\;e_2^{p_2}]^p)&=fv(e_1^{p_1})\cup fv(e_2^{p_2})\\
		fv([\mbox{let}\;y\;e_1^{p_1}\;e_2^{p_2}]^p)&=fv(e_1^{p_1})\cup fv(e_2^{p_2})\backslash\{y\}\\
		fv([\mbox{let rec}\;f\;e_1^{p_1}\;e_2^{p_2}]^p)&=fv(e_1^{p_1})\cup fv(e_2^{p_2})\backslash\{f\}\\
		fv([\mbox{case}\;e^{p'}\;(s_1,\cdots,s_n)\;(e_1^{p_1},\cdots,e_n^{p_n})]^p)&=fv(e^{p'})\cup fv(e_1^{p_1})\cup\cdots\cup fv(e_n^{p_n})\backslash(\tau(s_1)\cup\cdots\cup\tau(s_n))\\
		fv([\mbox{ref}\;e^{p'}]^p)&=fv(e^{p'})\\
		fv([!e^{p'}]^p)&=fv(e^{p'})\\
		fv([e_1^{p_1}\;:=\;e_2^{p_2}]^p)&=fv(e_1^{p_1})\cup fv(e_2^{p_2})\\
	\end{align*}
	where $\tau(s)$, for a pattern $s$, is denoted as:
	$$
	\tau(s)=
		\left\{\begin{matrix}
			\{x\} & \mbox{if}\;s=x\\ 
			\emptyset & \mbox{otherwise}
		\end{matrix}\right.
	$$
\end{definition}

\iffalse
\begin{definition}[Bound variables]\label{def:bv}
	The set of bound variables is given by:
	\begin{align*}
		bv(x^p)&=\emptyset\\
		bv(c^p)&=\emptyset\\
		bv([\lambda\;y.e_1^{p'}]^p)&=bv(e_1^{p'})\cup\{y\}\\
		bv([e_1^{p'}\;e_2^{p''}]^p)&=bv(e_1^{p'})\cup bv(e_2^{p''})\\
		bv([\mbox{let}\;y\;e_1^{p'}\;e_2^{p''}]^p)&=bv(e_1^{p'})\cup bv(e_2^{p''})\cup\{y\}\\
		bv([\mbox{let rec}\;f\;e_1^{p'}\;e_2^{p''}]^p)&=bv(e_1^{p'})\cup bv(e_2^{p''})\cup\{f\}\\
		bv([\mbox{case}\;e^{p'}\;\pi^{p''}]^p)&=bv(e_1^{p'})\cup bv(\pi)\\
		bv([(s\;e^{p'})\;\pi])&=bv(e^{p'})\cup bv(\pi)\cup\tau(s)\\
		bv([(s\;e^{p'})])&=bv(e^{p'})\cup\tau(s)\\
		bv([\mbox{ref}\;e^{p'}]^p)&=bv(e^{p'})\\
		bv([!e^{p'}]^p)&=bv(e^{p'})\\
		bv([e_1^{p'}\;:=\;e_2^{p''}]^p)&=bv(e_1^{p'})\cup bv(e_2^{p''})\\
	\end{align*}
\end{definition}
\fi

\subsection{Environments and stores}\label{sec:EnvSto}
We will now introduce the binding model used in the semantics, where we will present the environments and stores.
Since the language we focus on introduces mutability, through the referencing, this needs to be reflected in our bindings model.
Here, the referencing constructs can also be seen as how locations, or pointers, are created and handled, as such we introduce notion of stores to describe how they are bound.

Since this language is a $\lambda$-calculus, the environment keeps the bindings we currently know and as such the environment is a function from variables to values.
The set of values, \cat{Values}, is comprised by:
\begin{itemize}
	\item All constants are values.
	\item Locations are values.
	\item Closures, $\langle x,e^{p'},env\rangle$ are values.
	\item Recursive closures, $\langle x,f,e^{p''},env\rangle$, are values.
	\item Unit values, $()$, are values.
\end{itemize}

Where a value $v\in\cat{Values}$ is an expression given by the following formation rules:
$$v::=c\mid\loc\mid\langle x,e^{p'},env\rangle\mid\langle x,f,e^{p''},env\rangle\mid ()$$

\begin{definition}[]
	The set of all environments, \cat{Env}, is the set of partial functions from variables to values, given as:
	$$\cat{Env}=\cat{Var}\rightharpoonup\cat{Values}$$
\end{definition}
Where $env\in\cat{Env}$ denotes an arbitrary environment in \cat{Env}.

\begin{definition}[Update of environments]
	Let $env\in\cat{Env}$ be an environment.
	We write $env[x\mapsto v]$ to denote the environment $env'$ where:
	\begin{align*}
		env'(y)=
		\left\{\begin{matrix}
			env(y) & \mbox{if}\;y\neq x\\\	 
			v & \mbox{if}\;y=x
		\end{matrix}\right.
	\end{align*}
\end{definition}

We also introduce a function which, for a given value $v$, returns all variables that is bound to $v$.
\begin{definition}[inverse env]
	Let $v$ be a value and $env\in\cat{Env}$ be an environment, the inverse function $env^{-1}$ is then given as:
	$$env^{-1}(v)=\{x\in dom(env)\mid env(x)=v\}$$
\end{definition}

The store is a function that keeps the location bindings currently known.
We also introduce a placeholder $next$, that represents the next free location.

\begin{definition}[]
	The set of all stores, \cat{Sto}, is the set of partial functions from locations, and the $next$ pointer, to values, given as:
	$$\cat{Sto}=\cat{Loc}\cup\{next\}\rightharpoonup\cat{Values}$$
\end{definition}
Where $sto\in\cat{Sto}$ denotes an arbitrary store in \cat{Sto}.

\begin{definition}[Update of stores]
	Let $sto\in\cat{Sto}$ be a store.
	We write $sto[\loc\mapsto v]$ to denote the store $sto'$ where:
	\begin{align*}
		sto'(\loc_1)=
		\left\{\begin{matrix}
			env(\loc_1) & \mbox{if}\;\loc_1\neq \loc\\\	 
			v & \mbox{if}\;\loc_1=\loc
		\end{matrix}\right.
	\end{align*}
\end{definition}

We also assume the existence of a function $new:\cat{Loc}\rightarrow\cat{Loc}$, which takes a location and finds the next location.
This function is used on the location $next$ points to, to get a new free location, which is not already bound in our store.

\subsection{Dependencies}\label{sec:DepFunc}
The goal of the collection semantics is to collect the semantic dependencies as they appear in a computation. 
To this end, we use a dependency function that will tell us for each variable and location occurrence what other, previous occurrences they depend upon.

As such, we use the dependency function to model the semantic flow of dependencies in an occurrence, where we present and ordering between those occurrences to denote the flow.

\begin{definition}[Dependency function]\label{def:DepFunc}
	The set of dependency functions, $\cat{W}$, is a set of partial functions from location and variable occurrences to a pair of dependencies, such that:
	$$\cat{W}=\cat{Loc}^P\cup\cat{Var}^P\rightharpoonup\Pow{\cat{Loc}^P}\times\Pow{\cat{Var}^P}$$
\end{definition}
A lookup in a dependency function $w$ is for an element $u^p\in\cat{Loc}^P\cup\cat{Var}^P$, such that:
$$w(u^p)=(\{\loc_1^{p_1},\cdots,\loc_n^{p_n}\},\{x_1^{p'_1},\cdots,x_m^{p'_m}\})$$
This should be read as: a lookup of an occurrence $u^p$, a variable or location occurrence, returns a pair of location and variable occurrences.
We also denote the pair, retrieved from the dependency function, which we call a dependency pair such that $(L,V)$ contains a set of location occurrences $L=\{\loc_1^{p_1},\cdots,\loc_n^{p_n}\}$ 
and a set of variable occurrences $V=\{x_1^{p'_1},\cdots,x_m^{p'_m}\}$.

\begin{definition}[Update of dependency functions]\label{def:DepExt}
	Let $w\in\cat{W}$ be a dependency function and $u^p$ be either a variable or location occurrence.
	We write $w[u^p\mapsto(L,V)]$ to denote the dependency function $w'$ where:
	\begin{align*}
		w'(v^q)=
		\left\{\begin{matrix}
			w(v^q) & \mbox{if}\;v^q\neq u^p\\\	 
			(L,V) & \mbox{if}\;v^q=u^p
		\end{matrix}\right.
	\end{align*}
\end{definition}

\begin{example}[]\label{ex:dep}
	Consider the occurrence from \cref{ex:write}, where we can infer the following bindings for a dependency function $w_{ex}$ over this occurrence:
	$$w_{ex}=[x^2\mapsto(\emptyset,\emptyset),z^4\mapsto(\emptyset,\emptyset),y^9\mapsto(\emptyset,\{x^5\}),\loc^2\mapsto(\emptyset,\emptyset),\loc^8\mapsto(\emptyset,\{z^7\})]$$
	Where $\loc$ is the location created from the reference construct.
	Here, we can see that the variable bindings are distinct, an the location $\loc$ is bound multiple times, for the program points $2$ and $8$.

	If we want to read a variable or location in $w_{ex}$, we must also know for which program point since there can exists multiple bindings for the same variable or location.
\end{example}

By considering \cref{ex:dep}, we would like to read the information from the location, that $x$ is an alias to.
As it is visible from the occurrence in \cref{ex:write}, we know that we should read from $\loc^8$, since we wrote that reference at the program point $8$.
We can also see that from $w_{ex}$ alone it is not possible to know which occurrence to read, since there are no order defined between the bindings.
We then present the notion of ordering, as a binary relation over program points:

\begin{definition}[]\label{def:BinRel}
	Let \cat{P} be a set of program points in an occurrence.
	Then $\sqsubseteq$ is a binary relation of \cat{P}, such that:
	$$\sqsubseteq\subseteq\cat{P}\times\cat{P}$$
\end{definition}

Since we are interested in the ordering of the elements in a dependency function $w$, we will define an instantiation of \cref{def:BinRel}.
Since $w$, is a function from occurrences to a pair of occurrences, we first present a function for getting the program points from a set of occurrences:

\begin{definition}[Occurring program points]\label{def:OccPP}
	Let $O$ be a set of occurrences, then $points(O)$ is given by:
	$$points(O)=\{p\in\cat{P}\mid\exists e^p\in O\}$$
\end{definition}

With \cref{def:OccPP} defined, we present the instantiation of \cref{def:BinRel} over a dependency function $w$:

\begin{definition}[]\label{def:RelPoint}
	Let $w\in\cat{W}$ be a dependency function.
	Then $\sqsubseteq_w$ is given by:
	$$\sqsubseteq_w\subseteq\{(p,p')\mid p,p'\in(points(dom(w)\cup points(ran(w))))\}$$
\end{definition}

As the dependency function $w$ is a model of which occurrences an occurrence is dependent on, the relation on $w$ should also model the order a value is evaluated in, as such we define the partial order over a dependency function.

\begin{definition}[Partial order of $w$]
	Let $w\in\cat{W}$ be a dependency function and $\sqsubseteq_w$ be a binary relation over $w$.
	We say that $w$ is partial order if $\sqsubseteq_w$ is a partial order.
\end{definition}

\begin{example}[]\label{ex:depRel}
	Consider the example from \cref{ex:dep}, if we introduce a binary relation over the dependency function $w_{ex}$, such that:
	$$\sqsubseteq_{w_{ex}}=\{(2,4),(2,9),(5,9),(2,8),(8,2)\}$$
	From this ordering, it is easy to see the ordering of the elements.
	The ordering we present also respects the flow the occurrence from \cref{ex:write} would evaluate to.
	We then know that the dependencies for the reference (that $x$ is an alias to) is for the largest binding of $\loc$.
\end{example}

As presented in \cref{def:DepFunc} and \cref{def:RelPoint}, the dependency function and the binary relation are used to define the flow of information.
As illustrated by \cref{ex:depRel}, we need to lookup the greatest of $\sqsubseteq_w$.

We first present a generic function for the greatest binding of a relation $\sqsubseteq$ of program points.

\begin{definition}[Greatest binding]\label{def:GBind}
	Let $u$ be an element, either a variable or location, and $S$ be a set of occurrences, then $uf(u,S)$ is given by:
	$$uf(u,S)=\inf\{u^p\in S\mid u^q\in S.q\sqsubseteq p\}$$
\end{definition}

Based on \cref{def:GBind}, we can present an instantiation of the function for the dependency function $w$ and an order over $w$, $\sqsubseteq_w$:

\begin{definition}[]
	Let $w$ be a dependency function, $\sqsubseteq_w$ be an order over $w$, $u$ be an element, that is either a variable or location, then $uf_{\sqsubseteq_w}$ is given by:
	$$uf_{\sqsubseteq_w}(u,w)=\inf\{u^p\in dom(w)\mid u^q\in dom(w.q\sqsubseteq_w p\}$$
\end{definition}

\begin{example}[]\label{ex:deplookup}
	As a continuation of \cref{ex:depRel}, we can now lookup the greatest element for an element, e.g., a variable or location.
	As we were interested in finding the greatest bindings a location is bound to in $w_ex$, we can now use the function $uf_{\sqsubseteq_w}$:
	$$uf_{\sqsubseteq_{w_ex}}(\loc,w_ex)=\inf\{\loc^p\in dom(w)\mid \loc^q\in dom(w). q\sqsubseteq_{w_ex} p\}$$
	Where the set we get for $\loc$ are as follows: $\{\loc^2,\loc^8\}$.
	From this, we find the greatest element:
	$$\loc^7=\inf\{\loc^2,\loc^8\}$$
	As we can see, from the $uf_{w_ex}$ function, we got $\loc^8$ which were the occurrence we wanted.
\end{example}

\subsection{Collection semantics}\label{sec:sem}
We will now introduce the big-step semantics for our language and highlight some interesting transition rules.
In the big-step semantics, the transitions are of the from:
\begin{align*}
env\vdash\left\langle e^{p'},sto,(w,\sqsubseteq_w),p\right\rangle\rightarrow\left\langle v,sto',(w',\sqsubseteq_w'),(L,V),p''\right\rangle
\end{align*}
Where $env\in\cat{Env}$, $sto\in\cat{Sto}$, and $w\in\cat{W}$.
This should be read as: given the store $sto$, a dependency function $w$, A relation over $w$, and the previous program point $p$, the occurrence $e^{p'}$ evaluates to a value $v$, an updated store $sto'$, an updated dependency function $w'$, 
a relation over $w'$, the dependency pair $(L,V)$, and the program point $p''$ reached after evaluating $e^{p'}$, given the bindings in the environment $env$.

The transition system is given by:
$$((\cat{Occ}\cup\cat{Values})\times\cat{Store}\times(\cat{W}\times(\cat{P}\times\cat{P})\times\cat{P},\rightarrow,
\cat{Values}\times\cat{Store}\times(\cat{W}\times(\cat{P}\times\cat{P}))\times\Pow{\cat{Loc}^P\times\cat{Var}^P}\times\cat{P})$$
A highlight of the rules for $\rightarrow$ can be found in \cref{fig:ColSem}, the rest can be found in \cref{App:ColSem}.

\begin{description}
	\item[\runa{Const}] rule, for the occurrence $c^{p'}$, is the simplest rule, as it has no premises and does not have any side effects.
		As constants are evaluated to the constant value, no dependencies are used, i.e., no variable or location occurrences are used to evaluate a constant.

	\item[\runa{Var}] rule, for the occurrence $x^{p'}$, uses the environment to get the value $x$ is bound to and uses dependency function $w$ to get its dependencies.
		To lookup the dependencies, the function $uf_{\sqsubseteq_w}$ is used to get the greatest binding a variable is bound to, in respect to the ordering $\sqsubseteq_w$.
		Since the occurrence of $x$ is used, it is added to the set of variable occurrences we got from the lookup of the dependencies for $x$.

	\item[\runa{Let}] rule, for the occurrence $[\mbox{let}\;x\;e_1^{p_1}\;e_2^{p_2}]^{p'}$, creates a local binding that can be used in $e_2^{p_2}$.
		The \runa{Let} rule evaluate $e_1^{p_1}$, to get the value $v$, that $x$ will be bound to in the environment for $e_2^{p_2}$, and the dependencies used to evaluate $e_1^{p_1}$ is bound in the dependency function.
		As we reach the program point $p_1$ after evaluating $e_1^{p_1}$, and it is also the program point before evaluating $e_2^{p_2}$, the binding of $x$ in $w$ is to the program points $p_1$.	

	\item[\runa{Ref}] rule, for the occurrence $[\mbox{ref}\;e^{p'}]^{p''}$, creates a new location and binds it in the store $sto$, to the value evaluated from $e^{p'}$.
		The \runa{Ref} rule also binds the dependencies, from evaluating the body $e^{p'}$, in the dependency function $w$ at the program point $p''$.
		As the \runa{Ref} rule creates a location (where we get the location from the $next$ pointer), and binds it in $sto$.
		The environment is not updated as \runa{Ref} does not in itself give any alias information.
		To create an alias for a location, it should be bound to a variable using the \runa{Let} rule.

	\item[\runa{Ref-read}] rule, for the occurrence $[!e^{p_1}]^{p'}$, evaluates the body $e^{p_1}$ to a value, that must be a location $\loc$, and reads the value of $\loc$ in the store.
		The \runa{Ref-read} rule also makes a lookup for the dependencies $\loc$ is bound to in the dependency function $w$.
		As there could be multiple bindings for $\loc$, in $w$, at different program points, we use the $uf_{\sqsubseteq_{w'}}$ function to get greatest binding of $\loc$ with respect to the ordering $\sqsubseteq_{w'}$, 
		and we also add the location occurrence $\loc^{p'}$ to the set of locations.

	\item[\runa{Ref-write}] rule, for the occurrence $[e_1^{p_1}\;:=\;e_2^{p_2}]^{p'}$, evaluate $e_1^{p_1}$ to a location $\loc$ and $e_2^{p_2}$ to a value $v$, and binds $\loc$ in the store $sto$ to the value $v$.
		The dependency function is also updated with a new binding for $\loc$ at the program point $p'$.
\end{description}

\begin{figure}[H]
	\setlength\tabcolsep{8pt}
	\begin{tabular}{l}
		\input{sections/appendix/ColRules/const.tex}\\[0.7cm]
		\input{sections/appendix/ColRules/var.tex}\\[0.7cm]
		\input{sections/appendix/ColRules/let.tex}\\[0.7cm]
		\input{sections/appendix/ColRules/ref.tex}\\[0.7cm]
		\input{sections/appendix/ColRules/refread.tex}\\[0.7cm]
		\input{sections/appendix/ColRules/refwrite.tex}
	\end{tabular}
	\caption{Selected rules from the semantics}
	\label{fig:ColSem}
\end{figure}

\iffalse
\begin{example}[Data-flow for abstractions]\label{ex:DFAbs}
The following program creates a local binding to the identity function and applies it twice.

\begin{lstlisting}[language=Caml, mathescape=true]
(let x ($\lambda$ y.(y$^1$))$^2$ (let z (x$^3$ 1$^4$)$^5$ (x$^6$ 2$^7$)$^{8}$)$^{9}$)$^{10}$
\end{lstlisting}
The transition tree can be found in \cref{FigEx.Abs}.
In the transition tree, we see that $\psi$ is extended a couple of times, where we will take a look at a couple of interesting extensions to $\psi$
The first time we evaluate the abstraction body, $\psi$ is on the following form:

$$\psi_2=(w_2=[x^{2}\mapsto(\emptyset,\emptyset), y^{4}\mapsto(\emptyset,\emptyset)],\sqsubseteq_w^2=\emptyset)$$
Here, the lookup of the parameter $y$ is simple, as there are only one occurrence, where we then know that $inf_{\psi_2}(y)=4$.

The second time we evaluate the body of the abstraction, $\psi$ is on the following form:

$$\psi_3=(w_3=[x^{2}\mapsto(\emptyset,\emptyset), y^{4}\mapsto(\emptyset,\emptyset), y^{7}\mapsto(\emptyset,\emptyset)],\sqsubseteq_w^2=\{4,7\})$$
Here, we now have two bindings for the parameter $y$, but since we also know that there are an ordering for the two occurrences of $y$, we then know that the program point $7$ is evaluated after $4$, as such we know that $inf_{\psi_3}(y)=7$.
\end{example}

\begin{landscape}
\subfile{../examples/DFAbs.tex}
\end{landscape}
\fi

\newpage
\section{Type system for data-flow analysis}\label{sec:TypeSys}
This section will introduce the type system for data-flow analysis on the language presented in \cref{sec:lang}.
Similarly to the language, the type rules, types and other parts are based on \cite{DVNicky}.
The type system presented in this section is a type checker for local analysis of occurrences.
The type checker assign types, presented in \cref{sec:types}, to occurrences given the basis, which will be presented in \cref{sec:basis}, and using the type assignment, which is presented in \cref{sec:Judge}.
Since the language contains local information as bindings, and global information as locations, the type checker should reflect this.
Since locations are a semantic notation, we will present internal variables to represent locations in the type system, as $\nu x,\; \nu y\in\cat{IVar}$ where \cat{IVar} is the syntactic category for internal variables.

As references are not always bound to variables, as such the reference does not contain any alias information, the analysis provides alias information used for evaluating an occurrence.
Here, we are going to introduce the basis for aliasing, as a partition of all variables and internal variables used in an occurrence.
As such, it is possible to analyse, from the type information of occurrences, which aliases are actually used.

We also impose some restrictions on the type system, where the first restrictions is that references cannot be bound to abstractions.
Since we do not introduce polymorphism, the use case for abstractions are reduced, as an abstraction cannot be used at multiple places.
Consider the following occurrence:
\begin{lstlisting}[language=Caml, mathescape=true]
(let x ($\lambda$ y.y$^1$)$^2$ (x$^3$ (x$^4$ 1$^5$)$^6$)$^7$)$^8$
\end{lstlisting}
To type the occurrences used in both places where we apply the abstractions, we type of the argument in the innermost application is empty, as it applies a constant.
For the second, and outermost, application, the argument type must contain the occurrence $x^4$, as it were used to evaluate the value for the argument.

\subsection{Types}\label{sec:types}
We denote the set of types as \cat{Types}, which are given by the following formation rules:

$$T::=(\delta,\kappa)\mid T_1 \rightarrow T_2$$

Here, we introduce two types, the base type $(\delta,\kappa)$ and the abstraction type $T_1 \rightarrow T_2$.
The idea is that an occurrence have the abstraction type if it represents an abstraction that takes an argument of type $T_1$ and returns a type of $T_2$.
The base type represent all other values, where $\delta$ represent the set of occurrences, used to evaluate an occurrence, and $\kappa$ represent the set of alias information.
Here, if an occurrence have a type containing alias information, then it represent a location, where if the occurrence have a base type with alias information, then the occurrence must represents a reference. 
If the occurrence have the abstraction type where either $T_1$ or $T_2$ have are base types with alias information, then the abstraction either takes a reference as input or returns a reference.

\begin{example}[]
Consider the following occurrence:
\begin{lstlisting}[language=Caml, mathescape=true]
(let x (3$^1$)$^2$ (let y (ref x$^3$)$^4$ (!y)))
\end{lstlisting}
Here, we can type $x$ with $(\emptyset,\emptyset)$ as $x$ is bound to a constant and there a no variables or internal variables used.
$y$ can then be given the type $(\emptyset,\{x,\nu y\})$, as the reference construct $ref$ creates a new reference, which $y$ is then an alias to, e.g., $y$ is bound to a location.
Her $\nu y$ represents the reference from $ref$, and can thus be given the type $(\{x^3\},\emptyset)$, where $\nu y$ is bound to a constant, because of $x$, but the occurrence $x^3$ were used, so it should be part of the set of occurrences $\delta$.
\end{example}

Since the type system approximates the occurrences used to evaluate an occurrence, we introduce two unions.
The first union is a simple union that expects the types to be similar, that is, only the base types are allowed to be different.
\begin{definition}[Type union]
	Let $T_1$ and $T_2$ be two types, then the type union, $\cup$, are as follows:
	\begin{equation*}
		T_1\cup T_2=
		\left\{\begin{matrix}
			\mbox{If } \; T_1=(\delta,\kappa) \;\mbox{and}\; T_2=(\delta',\kappa')  & \mbox{then} \; (\delta\cup\delta',\kappa\cup\kappa')\\
			\mbox{else if } \; T_1=T_1'\rightarrow T_1''\;\mbox{and}\;T_2=T_2'\rightarrow T_2'' & \mbox{then} \; (T_1'\cup T_2')\rightarrow (T_1''\cup T_2'')
		\end{matrix}\right.
	\end{equation*}
\end{definition}

The second type union, is to add additional type information to an arbitrary type.
This type union is used to add an occurrence to a type, e.g., in the \runa{Var} rule where the variable occurrence needs to be added to the type of that variable.
\begin{definition}[Base type union]
	Let $T$ be an type and $(\delta,\kappa)$ be a base type, then the union of these are as follows:
	\begin{equation*}
		T\sqcup (\delta,\kappa)=
		\left\{\begin{matrix}
			\mbox{If } \; T=(\delta',\kappa')  & \mbox{then} \; (\delta\cup\delta',\kappa\cup\kappa')\\
			\mbox{else if } \; T=T_1\rightarrow T_2 & \mbox{then} \; T_1\rightarrow(T_2\sqcup (\delta,\kappa))
		\end{matrix}\right.
	\end{equation*}
\end{definition}

\subsection{Basis and type environment}\label{sec:basis}
Next, we will present the basis and type environment for the type system.
The basis we are presenting here are assumptions used by the type checker, in addition to the assignment of types which are presented in \cref{sec:Judge}, where we are going to present a type base for aliasing and an approximated order of program points.

We will also introduce the type environment, which are similar to the environment and store used in the semantics, as the type environment keeps track of the type of variables and internal variables.
As such, the type environment is also a approximation of the dependency function, as the purpose of the type system is to collect information about which occurrences are used and what alias information is used.

Similar to the lookup of the greatest binding for the dependency function, we are going to introduce an instantiation of the function from \cref{def:GBind} for the type environment in respect to the basis for approximated order of program points.
\bigskip

We will then introduce the type base for aliasing, as a partition of variables and internal variables used in an occurrence.

\begin{definition}[Type Base for aliasing]
	For an occurrence $o$, let $var$ be the set of all variables and $ivar$ be the set of all internal variables in $o$.
	The type base $\kappa^0=\{\kappa^0_1,\cdots,\kappa^0_n\}$ is then a partition of $var\cup ivar$, where $\kappa_i^0\cap\kappa_j^0=\emptyset$ for all $i\neq j$.
\end{definition}

The idea behind the base for type alias $\kappa_0$ is to make a partition of the variables and internal variables used in an occurrence.
This partition represents the assumption about which variables are actually an alias to internal variables.
As such multiple variables can only belong to the same element $\kappa_0^i\in\kappa_0$, if there also exists an internal variable in $\kappa_0^i$.

\begin{definition}[Approximated order of program points]
	An approximated order of program points $\Pi$ is a pair, such that: 
	$$\Pi=(\cat{P},\sqsubseteq_\Pi)$$
	where
	\begin{itemize}
		\item \cat{P} is the set of program points in an occurrence,
		\item $\sqsubseteq_\Pi\subseteq\cat{P}\times\cat{P}$, where
	\end{itemize}
\end{definition}

The approximated order of program points is an assumption about the order for program points for an occurrence $o$, as such, this approximation should be an approximation of the order that that can be derived from the semantics, presented in \cref{sec:sem}, for $o$.

\begin{definition}[Partial order of $\Pi$]
	Let $\Pi=(\cat{P},\sqsubseteq_\Pi)$ be an approximated order of program points.
	We say that $\Pi$ is a partial order if $\sqsubseteq_\Pi$ is a partial order.
\end{definition}
\bigskip

Next, we will introduce the type environment:
\begin{definition}[Type Environment]
	A type environment $\Gamma$ is a partial function $\Gamma:\cat{Var}^P\cup\cat{IVar}^P\rightharpoonup\cat{Types}$
\end{definition}

\begin{definition}[Updating Type Environments]
	Let $\Gamma$ be a type environment.
	We write $\Gamma[u^p:T]$, for an occurrence $u^p$, to denote the type environment $\Gamma'$ where:
	\begin{align*}
		\Gamma'(y^{p'})=
		\left\{\begin{matrix}
			\Gamma(y^{p'}) & \mbox{if}\;y^{p'}\neq u^{p}\\\	 
			T & \mbox{if}\;y^{p'}=u^{p}
		\end{matrix}\right.
	\end{align*}
\end{definition}

Similar to the lookup of dependencies in the semantics, we need to similarly define how to lookup in the type environment.
As the type environment contains both local information, for local declarations, and global information, for references, both cases should be handled.

For local information we introduce, similarly to lookup in the dependency function, and instantiation of the function presented in \cref{def:GBind}.
The lookup is for information in the type environment, over the relation between program points defined by the basis for approximated order of program points.

\begin{definition}[]\label{def:GBindPi}
	Let $u\in \cat{Var}\cup\cat{IVar}$, be either a variable or internal variable, $\Gamma$ be a type environment, and $\Pi$ be the approximated order of program points that is a partial order, then $uf_{\sqsubseteq_\Pi}$ is given by:
	$$uf_{\sqsubseteq_\Pi}(u,\Gamma)=\inf\{u^p\in dom(\Gamma)\mid u^q\in dom(\Gamma).q\sqsubseteq_\Pi p\}$$
\end{definition}

Where the lookup for global information needs to be handled differently as the language contains pattern matching, and as such, the language can contain different path of evaluation (where each pattern in the pattern matching construct introduces a new path).
To handle the lookup of global information, we will first introduce the notion of $p$-chains as chains of program points with respect to the approximated order of program points, where the maximal program point is $p$.
The idea of these $p$-chains is to describe the history behind an occurrence $u^p$, and can thus be used to describe what an internal variable depends on.

\begin{definition}[$p$-chains]
	Let $\Pi$ be an approximated order of program points, that is a partial order, and $p$ be a program point.
	We then say that a $p$-chain, denoted as $\Pi_p^{*}$, is a maximal chain of with the maximal element $p$ with the respect to the order $\Pi$.
	As such, any $p$-chain is a total order, where $\Pi_p^{*}$ does not contain any pairs $(p,q)\in\sqsubseteq_\Pi$, where $p\neq q$, then $(p,q)\notin\sqsubseteq_{\Pi_p^{*}}$.
\end{definition}

We also denote $\Pi_p^{*}\in\Pi$, if the $p$-chain $\Pi_p^{*}$ can be be derived from $\Pi$.
Since there can exists multiple paths in an occurrence, we define the set of all $p$-chains as follows:

\begin{definition}[]
	Let $\Pi$ be an approximated order of program points and $p$ be a program point.
	We say that $\Upsilon_p$ is the set of all $p$-chains in $\Pi$.
\end{definition}

Since $\Upsilon_p$ contains all $p$-chains in an approximated order of program points $\Pi$, with $p$ as the maximal element, we can then define the function to lookup all greatest element less than or equal to $p$.

\begin{definition}[]\label{def:GBindUps}
	Let $u\in \cat{Var}\cup\cat{IVar}$, be either a variable or internal variable, $\Gamma$ be a type environment, and $\Upsilon_p$ be a set of $p$-chains, then $uf_{\Upsilon_p}$ is given by:
	$$uf_{\Upsilon_p}(u,\Gamma)=\bigcup_{\Pi_p^{*}\in\Upsilon_p}uf_{\Pi_p^{*}}(u,\Gamma)$$
\end{definition}

The function, defined in \cref{def:GBindUps}, takes the union of the greatest binding, for an element, for each $p$-chain using the function defined in \cref{def:GBindPi}.

\subsection{The type system}\label{sec:Judge}
We will now present the judgement and type rules for the language, that is, how we assign types to occurrences.

The type judgement is defined as:
$$\Gamma,\Pi\vdash e^p: T$$
And should be read as: the occurrence $e^p$ has type $T$, given the dependency bindings $\Gamma$ and the approximated order of program points $\Pi$.

A highlight of type rules can be found in \cref{fig:TypeSys}, and all type rules can be found in \cref{App:TypeSys}.

\begin{description}
	\item[\runa{T-Const}] rule, for occurrence $c^p$, is the simplest type rule, as there is nothing to track for constants, and as such it has the type $(\emptyset,\emptyset)$.

	\item[\runa{T-Var}] rule, for occurrence $x^p$, looks up the type for $x$ in the type environment, by finding the greatest binding using \cref{def:GBindPi}, and add the occurrence $x^p$ to the type.

	\item[\runa{T-Let-1}] rule, for occurrence $[\mbox{let}\;x\;e_1^{p_1}\;e_2^{p_2}]^p$, creates a local binding for a variable, with the type of $e_1^{p_1}$ that can be used in $e_2^{p_2}$.
		The \runa{T-Let-1} rule assumes that the type of $e_1^{p_1}$ is a base type with alias information, i.e., $\kappa\neq\emptyset$.
		If this is the case, then $e_1^{p_1}$ must evaluate to a location, in the semantics.
		The other cases, when $e_1^{p_1}$ is not a base type with alias information, are handled by the \runa{T-Let-2} rule.
		Since a pattern can be a variable, we updates the type environment with the type of $e^p$.

	\item[\runa{T-Case}] rule, for occurrence $[\mbox{case}\;e^{p}\;\tilde{\pi}\;\tilde{o}]^{p'}$, is an over-approximation of all cases in the pattern matching expression, by taking an union of the type of each case.
		Since the type of $e^p$ is used to evaluate the pattern matching, we also add this type to the type of the pattern matching.

	\item[\runa{T-Ref-read}] rule, for occurrence $[!e^{p}]^{p'}$, is used to retrieve the type of references, where $e^p$ must be a base type with alias information.
		Since the type system is an over-approximation, there can be multiple internal variables in $\kappa$ and multiple occurrences we need to read from.
		As such we need to lookup for all internal variables and also possible for multiple program points.
		As such, we use the $uf_{Upsilon_{p'}}$ to lookup for all $p'$-chains.
\end{description}

\begin{example}[Data-flow for abstractions]
	Consider the following occurrence for application:
	\begin{lstlisting}[language=Caml, mathescape=true]
		(($\lambda$ y.(PLUS 3$^1$ y$^2$)$^3$)$^4$ 5$^5$)$^6$
	\end{lstlisting}
	The derivation tree for the occurrence can be found in \cref{FigEx.TAbs}.
	Here, we show two applications, for \runa{T-App} and \runa{T-App-const}, where we create an abstraction that adds the constant $3$ to the argument of the abstraction, and applying the constant $5$ to the abstraction.

	When typing the abstraction, we need too make an assumption about the parameter $y$ and the body.
	As we are applying a constant to the argument, we can make an assumption that the type of the parameter should be $(\emptyset,\emptyset)$.

	Based on this assumption for the type, we can then type the body of the abstraction.
	As the body is an application for a functional constant, \runa{T-App-const}, we take a union for the types of each argument.
\end{example}

\begin{example}[Data-flow for references]
	Consider the following occurrence:
	\begin{lstlisting}[language=Caml, mathescape=true]
		(let x (ref 1$^1$)$^2$ (let y (x$^3$) (!x$^4$)$^5$)$^6$)$^7$
	\end{lstlisting}
	The derivation tree for the occurrence can be found in \cref{FigEx.TRef}.
	Here, we show the typing of references where we create a reference and create 2 aliases for it before reading from the reference.
	When typing the reference, it modifies the base type $\Gamma$ with a new internal variable.
	From the type information, it is clear that only $x$ and the internal variable $\nu x$ is used.
\end{example}

\begin{figure}[H]
	\setlength\tabcolsep{8pt}
	\begin{tabular}{l}
		\input{sections/appendix/TSRules/const.tex}\\[0.7cm]
		\input{sections/appendix/TSRules/var.tex}\\[0.7cm]
		\input{sections/appendix/TSRules/let1.tex}\\[0.7cm]
		\input{sections/appendix/TSRules/case.tex}\\[0.7cm]
		\input{sections/appendix/TSRules/refread.tex}\\[0.7cm]
	\end{tabular}
	\caption{Selected rules from the type system}
	\label{fig:TypeSys}
\end{figure}

\begin{landscape}
\begin{figure}
\begin{tabular}{l}
\inference[\runa{T-Let}]
{
	\inference[\runa{T-Abs}]
	{
		\inference[\runa{T-App-const}]
		{
			\inference[\runa{T-Const}]{}
			{\Gamma',\Pi\vdash 3^1:(\emptyset,\emptyset)}
			\;\;
			\inference[\runa{T-Var}]{}
			{\Gamma',\Pi\vdash y^2:(\{y^2\},\emptyset)}
		}
		{\Gamma',\Pi\vdash [PLUS\;3^1\;y^2]^3:(\{y^2\},\emptyset)}
	}
	{\Gamma,\Pi\vdash [\lambda\;y.(PLUS\;3^1\;y^2)^3]^4:(\emptyset,\emptyset)\rightarrow(\{y^2\},\emptyset)}
	\;\;
	\inference[\runa{T-Const}]{}
	{\Gamma,\Pi\vdash[5^5]^8:(\emptyset,\emptyset)}
}
{\Gamma,\Pi\vdash[(\lambda\;y.(PLUS\;3^1\;y^2)^3)^4\;5^5]^6:(\{y^2\},\emptyset)}\\[0.4cm]
Where $\Gamma'=\Gamma[y^{p_0}:(\emptyset,\emptyset)]$
\end{tabular}
\caption{Abstraction type example}\label{FigEx.TAbs}
\end{figure}

\begin{figure}
\begin{tabular}{l}
\inference[\runa{T-Let}]
{
	\inference[\runa{T-Ref}]
	{
		\inference[\runa{T-Const}]{}
		{\Gamma,\Pi\vdash 1^1:(\emptyset,\emptyset)}
	}
	{\Gamma,\Pi\vdash[ref\;1^1]^2:(\emptyset,\{\nu x\})}
	\;\;
	\inference[\runa{T-Let}]
	{
		\inference[\runa{T-Var}]{}
		{\Gamma',\Pi\vdash x^3:(\{x^3\},\{\nu x\})}
		\;\;
		\inference[\runa{T-Ref-read}]
		{
			\inference[\runa{T-Var}]{}
			{\Gamma'',\Pi\vdash x^4:(\{x^4,\nu x^5\},\{\nu x\})}
		}
		{\Gamma'',\Pi\vdash [!x^4]^5:(\{x^4,\nu x^5\},\emptyset)}
	}
	{\Gamma',\Pi\vdash [let\;y\;(x^3)\;(!x^4)^5]^6:(\{x^4,\nu x^5\},\emptyset)}
}
{\Gamma,\Pi\vdash[let\;x\;(ref\;1^1)^2\;(let\;y\;(x^3)\;(!x^4)^5)^6]^7:(\{x^4,\nu x^5\},\emptyset)}\\[0.4cm]
Where $\Gamma=\Gamma[\nu x^2\mapsto(\emptyset,\emptyset)]$, $\Gamma'=\Gamma[x^2\mapsto(\emptyset,\{\nu x\})]$, and $\Gamma''=\Gamma'[y^3\mapsto(\{x^3\},\{\nu x\})]$
\end{tabular}
\caption{Reference type example}\label{FigEx.TRef}
\end{figure}

\end{landscape}

\section{Soundness}\label{sec:Soundness}
We will now show the soundness of the type system, i.e., the type of an occurrence correspond to the dependencies and the alias information from the semantics.
To show that the type system is sound, we will first introduce the type rules for values and the relation between the semantics and the type system.
After that, we will present some properties in the semantics and type system that are used in the soundness proof.
And lastly, we will show the soundness of the type system.

\subsection{Type rules for values}
For the sake of proving the type system, we present type rules for the values presented int \cref{sec:EnvSto}, where the type rules is given in \cref{fig:ValTypeRules}.

As the values for closures and recursive closures contains an environment, from where they where declared, as such, before introducing the type rules for values we will present the notion for well-typed environments.

\begin{definition}[Environment judgement]\label{def:TEnv}
	Let $v_1,\cdots,v_n$ be values such that $\Gamma,\Pi\vdash v_i:T_i$, for $1\leq i\leq n$.
	Let $env$ be an environment given by $env=[x_1\mapsto v_1,\cdots,x_n\mapsto v_n]$, $\Gamma$ be a type environment, and $\Pi$ be the approximated order of program points.
	We say that:
	$$\Gamma,\Pi\vdash env$$
	iff 
	\begin{itemize}
		\item For all $x_i\in dom(env)$ then $\exists x_i^p\in dom(\Gamma)$ where $\Gamma(x_i^p)=T_i$ then 
			$$\Gamma,\Pi\vdash env(x_i):T_i$$
	\end{itemize}
\end{definition}

In \cref{def:TEnv} we show the notion of well-typed environments, which states that: given the type of all values, $T_i$, for all variables, $x_i$, bound in the $env$, 
then there exists an occurrence of $x$ in $\Gamma$, where the type from looking up for that occurrences is is $T_i$.
As such, we know that all bindings $x$ in $env$ also have a type in $\Gamma$ from when $x$ were declared.

\begin{description}
	\item[\runa{Constant}] type rule differs from the rule \runa{T-Const}, since most occurrences can evaluate to a constant and as such we know that its type should be a base type.
		Since constants can depend on other occurrences, we know that $\delta$ can be non-empty, but since constants are not locations, we also know that it cannot contain alias information, and as such $\kappa$ should be empty.

	\item[\runa{Location}] type rule represents locations, where we know that it must be a base type.
		Since locations can depend on other occurrences, we know that $\delta$ can be non-empty.
		As locations can contains alias information, and that a location is considered to always be an alias to itself, we know that $\kappa$ can never be empty, as it should always contain an internal variable. 

	\item[\runa{Closure}] type rule represents abstraction, and as such we know that it should have the abstraction type, $T_1\rightarrow T_2$, where we need to make an assumption about the argument type $T_1$.
		Since a closure contains the parameter, body, and the environment for an abstraction from when it were declared, we also need to handle those part in the type rule.
		The components of the closure is handled in the premises, where the environment must be well-typed.
		We also type the body of the abstraction, where we know that we need to update the type environment with the type $T_1$ for its parameter, Where we type the body with $T_2$.

	\item[\runa{Recursive closure}] type rules is similar to the \runa{Closure} rule, but since this is a recursive closure, we additionally need to update the type environment with the name of the recursive binding to the type of the abstraction.

	\item[\runa{Unit}] type rule simply have the base type, as it is not an abstraction and it also cannot have alias information.
		As the unit value is introduced from writing to references, $o=[o_1\;:=\;o_2]^p$, we know that from the type rule \runa{Ref-write} that the dependencies from the occurrence $o$ should also contain the set of occurrences.
		As such, the \runa{Unit} rule also contains a set of occurrences, $\delta$.
\end{description}

\begin{figure}[H]
	\setlength\tabcolsep{8pt}
	\begin{tabular}{l}
		\runa{Constant}\\[0.2cm]
			\inference[]{}
				{\Gamma,\Pi\vdash  c:(\delta, \emptyset)}\\[1cm]

		\runa{Location}\\[0.2cm]
			\inference[]{}
				{\Gamma,\Pi\vdash  \loc:(\delta, \kappa)}\\
				Where $\kappa\neq\emptyset$\\[1cm]

		\runa{Closure}\\[0.2cm]
			\inference[]
				{
					\Gamma,\Pi\vdash env \\
					\Gamma[x^{p}:T_1],\Pi\vdash e^{p'}:T_2
				}
				{\Gamma,\Pi\vdash \left\langle x^{p}, e^{p'}, env \right\rangle^{p''}:T_1\rightarrow T_2}\\[1cm]

		\runa{Recursive closure}\\[0.2cm]
			\inference[]
				{
					\Gamma,\Pi\vdash env \\
					\Gamma[x^{p}:T_1,f^{p'}:T_1\rightarrow T_2],\Pi\vdash e^{p''}:T_2
				}
				{\Gamma,\Pi\vdash \left\langle x^{p}, f^{p'}, e^{p''}, env \right\rangle^{p_3}:T_1\rightarrow T_2}\\[1cm]

		\runa{Unit}\\[0.2cm]
			\inference[]{}
				{\Gamma,\Pi\vdash  ():(\delta,\emptyset)}\\[0.5cm]
	\end{tabular}
	\caption{Type rules for values}
	\label{fig:ValTypeRules}
\end{figure}

\subsection{Agreement}
This section introduces the agreement between the type system and the semantics, where we will present the relation between the binding models in the type system and semantics, and show the relation between them.

We will first introduce the agreement between the binding models, i.e., show how the type environment and approximated order of program points relate to the environment, store, and dependency function.
Then we will show the type agreement, i.e., show the conditions for when a type agrees with the semantics.
As such, the type agreement needs to show when the dependencies agrees and alias if the alias information agrees with the basis.
\bigskip

The first agreement we present is the environment agreement, which ensures that that the type environment and approximated order of program points are a good approximation of the binding model in the semantics, i.e., 
for the environment $env$, store $sto$, dependency function $w$, and the relation of program points over $w$.

Here $env$, $sto$, and $w$ contains information for en evaluation in the semantics, either before or after an evaluation.
The type environment $\Gamma$ contains the local information for variable bindings and global information for internal variables, and the approximated order of program points $\Pi$ is an approximation of all program points in an occurrence.

\begin{definition}[Environment agreement]\label{def:EnvAgree}
	Let $(w,\sqsubseteq_w)$ be a pair containing the dependency function and a relation over it, $env$ be an environment, $sto$ be the a store, $\Gamma$ be a type environment, and $\Pi$ be an approximated program point order.
	We say that:
	$$(env,sto,(w,\sqsubseteq_w))\models(\Gamma,\Pi)$$
	if 
	\begin{enumerate}
		\item $\forall x\in dom(env).(\exists x^p\in dom(w))\wedge(x^p\in dom(w)\Rightarrow \exists x^p\in dom(\Gamma))$
		\item $\forall x^p\in dom(w).x^p\in dom(\Gamma)\Rightarrow env(x)=v\wedge w(x^p)=(L,V)\wedge\Gamma(x^p)=T.\\(env,v,(w,\sqsubseteq_w),(L,V))\models (\Gamma,T)$
		\item $\forall \loc\in dom(sto).(\exists \loc^p\in dom(w))\wedge(\exists \nu x.\forall p\in\{p'\mid\loc^{p'}\in dom(w)\}\Rightarrow\nu x^p\in dom(\Gamma))$
		\item $\forall \loc^p \in dom(w).\exists\nu x^{p}\in dom(\Gamma)\Rightarrow w(\loc^p)=(L,V)\wedge\Gamma(\nu x^{p})=T.(env,\loc,(w,\sqsubseteq_w),(L,V))\models T$
		\item if $p_1\sqsubseteq_w p_2$ then $p_1\sqsubseteq_\Pi p_2$
	\item $\forall \loc^p\in dom(w).\exists \nu x^p\in dom(\Gamma)\Rightarrow\exists p'\in\cat{P}.uf_{\sqsubseteq_w}(\loc,w)\in uf_{\Upsilon_{p'}}(\nu x,\Gamma)$
	\end{enumerate}
\end{definition}
The idea behind the environment agreement is that we need to make sure that semantics and type system talks about the same, i.e., if the dependencies in the semantics is also captured in the type environment, the alias information is captured, 
that $\Pi$ is a good approximation, in respect to $w$, and the $p$-chains captures the global occurrence.
As such, the type environment focuses on three areas: \cat{1)} local information variables, \cat{2)} the global information for references, and \cat{3)} the approximated order of program points.
It should be noted at the agreement only relates the information known by $env$, $sto$, and $w$.

\begin{description}
	\item[1)] The agreement for local information only relates the information currently known by $env$, and that the information known by $w$ and $\Gamma$ agrees, in respect to \cref{def:TAgree}.
		This is ensured by \cat{1)} and \cat{2)}.

	\item[2)] We similarly handles agreement for the global information known, which is ensured by \cat{3)} and \cat{4)}.
		Since $\Gamma$ contains the global information for references, we require that there exists a corresponding internal variable to the currently known locations, by comparing them by program points.
		We also ensures that the dependencies from a location occurrence agrees with the type of a corresponding internal variable occurrence, in respect to \cref{def:TAgree}.

	\item[3)] We also needs to ensure that $\Pi$ is a good approximation of the order $\sqsubseteq_w$ and the greatest binding function for $p$-chains ensures that we always get the necessary reference occurrences.
	\cat{5)} ensures that if an order is defined in $\sqsubseteq_w$, then $\Pi$ also agrees on this order.

	For \cat{6)}, we need to ensure that for any location currently known the exists a corresponding internal variable where, getting the greatest binding of this occurrence, $\loc^p$, then there exists a program point $p'$, 
	such that looking up all greatest bindings for the $p'$-chain, there exists an internal variable occurrence that corresponds to $\loc^p$.
\end{description}
\bigskip

With the environment agreement defined, we can present the type agreement.
As the type can be abstractions and base types, with or without alias information, we have different requirements for handling them, as such we relate each requirement to a value
Here, the idea is that if the value is a location, then we check that both the set of occurrences agrees with the dependency pair, presented in \cref{def:DepAgree}, 
and check if the alias information agrees with the semantics, \cref{def:AliasAgree}.
If the value is not a location, then the type can either be an abstraction type or base type.
For the base type, we check that the agreement between the set of occurrences and the dependency pair agrees.
If the type is an abstraction, then we check that $T_2$ agrees with binding model.
We are only concerned about the return type $T_2$ for abstractions, since if the argument parameter is used in the body of the abstraction, then the dependencies would already be part of the return type.

\begin{definition}[Type agreement]\label{def:TAgree}
	Let $env$ be an environment, $w$ be a dependency function, $\sqsubseteq_w$ be a relation over $w$, $(L,V)$ be a dependency pair, and $T$ be a type.
	We say that:
	$$(env,v,(w,\sqsubseteq_w),(L,V))\models(\Gamma,T)$$
	iff
	\begin{itemize}
		\item $v\neq\loc$ and $T=T_1\rightarrow T_2$:
		\begin{itemize}
			\item $(env,v,(w,\sqsubseteq_w),(L,V))\models(\Gamma,T_2)$
		\end{itemize}

		\item $v\neq\loc$ and $T=(\delta,\kappa)$:
		\begin{itemize}
			\item $(env,(L,V))\models\delta$
		\end{itemize}

		\item $v=\loc$ then $T=(\delta,\kappa)$ where:
		\begin{itemize}
			\item $(env,(L,V))\models\delta$
			\item $(env,(w,\sqsubseteq_w),v)\models(\Gamma,\kappa)$
		\end{itemize}
	\end{itemize}
\end{definition}

\begin{definition}[Dependency agreement]\label{def:DepAgree}
	Let $env$ be an environment, $(L,V)$ be a dependency pair, and $\delta$ be a set of occurrences.
	We say that:
	$$(env,(L,V))\models\delta$$
	if
	\begin{itemize}
		\item $V\subseteq\delta$,
		\item For all $\loc^p\in L$ where $env^{\loc}\neq\emptyset$, we then have $\{x\in dom(env)\mid env(x)=\loc\}\subseteq \kappa_i^0$ for a $\kappa_i^0\in\delta$
		\item For all $\loc^p\in L$ where $env^{\loc}=\emptyset$ then there exists a $\kappa_i^0\in\delta$ such that $\kappa_i^0\subseteq\cat{IVar}$
	\end{itemize}
\end{definition}

The dependency agreement, defined in \cref{def:DepAgree}, ensures that $\delta$ at leas contains all information from the dependency pair.

\begin{definition}[Alias agreement]\label{def:AliasAgree}
	Let $env$ be an environment, $w$ be a pair of a dependency function, $\sqsubseteq_w$ be a relation over $w$, $\loc$ be a location, and $\kappa$ be an alias set.
	We say that:
	$$(env,(w,\sqsubseteq_w),\loc)\models(\Gamma,\kappa)$$
	if
	\begin{itemize}
		\item $\exists \loc^p\in dom(w).\nu x^p\in dom(\Gamma)\Rightarrow\nu x\in\kappa$
		\item $env^{-1}(\loc)\neq\emptyset.\exists \kappa^0_i\in\kappa^0\Rightarrow
			(env^{-1}(\loc)\subseteq\kappa^0_i)\wedge(\exists \loc^p\in dom(w).\nu x^p\in dom(\Gamma)\Rightarrow\\\nu x\in\kappa^0_i\wedge\nu x\in\kappa)$
		\item $env^{-1}(\loc)=\emptyset.\exists \kappa^0_i\in\kappa^0\Rightarrow
			(\exists \loc^p\in dom(w).\nu x^p\in dom(\Gamma)\Rightarrow\nu x\in\kappa^0_i\wedge\nu x\in\kappa)$
	\end{itemize}
\end{definition}

The alias agreement, defined in \cref{def:AliasAgree}, ensures that the alias information in $\kappa$ is also known the environment.
To do this, we ensure that if there exists alias information in the environment $env$, then there exists an alias base $\kappa^0_i\in\kappa^0$ such that the currently know alias information known in 
in $env$ is a subset of $\kappa^0_i$, and that there exists a $\nu x\in\kappa$, such that $\nu x\in \kappa^0_i$.
If there is no currently known alias information, we simply check that there exists a corresponding internal variable, that is part of an alias base.

\subsection{Properties}
Before we present the soundness proof, we will first present some properties about the semantics and type system.
The first property we present is for the dependency function,since the dependency function is global, and as such they can contain side effects after an evaluation.
This property states that if any new variable bindings is introduced to the dependency function, by evaluating an occurrence $e^p$, those variables are not free in $e^p$.

\begin{lemma}[History]\label{lemma:His}
	Suppose $e^p$ is an occurrence, that
	$$env\vdash\left\langle e^{p},sto,(w,\sqsubseteq_w),p'\right\rangle\rightarrow\left\langle v,sto',(w',\sqsubseteq_w'),(L,V),p''\right\rangle$$
		and $x^{p_1}\in dom(w')\backslash dom(w)$.
		Then $x\notin fv(e^{p})$
\end{lemma}
The proof for \cref{lemma:His} can be found in \cref{app:HisProof}.

%\subfile{HisProof/index.tex}

The second property is the strengthening of the type environment, which states that if there is a binding the type environment, used to type an occurrence $e^p$, and the variables is not free in $e^p$ then the binding can be removed.

\begin{lemma}[Strengthening]\label{lemma:Strength}
	If $\Gamma[x^{p'}:T'],\Pi\vdash e^{p}:T$ and $x\notin fv(e^p)$, then $\Gamma,\Pi\vdash e^{p}:T$
\end{lemma}
The proof for \cref{lemma:Strength} can be found in \cref{app:StrProof}.
%\subfile{StrProof/index.tex}

With history, \cref{lemma:His}, and strengthening, \cref{lemma:Strength}, defined we can then present the soundness theorem.
This theorem compares the semantics, for an occurrence, to a type rule that concludes this occurrence.
Since we are interested in, if the type system is a sound approximation of the semantics, we need to make sure that an evaluation of an occurrence, and the type for the occurrence agrees.
As such, we assume that the type environment and approximated order of program points are in an agreement with the binding models in the semantics, and we also assume that the environment is well-typed.

Based on these assumptions, we then need to make sure that, after an evaluation, we are still in agreement, we can type the value, and the type is in agreement with the semantics.

\begin{theorem}[Soundness of type system]
	Suppose $e^{p'}$ is an occurrence where
	\begin{itemize}
		\item $env\vdash\left\langle e^{p'},sto,(w,\sqsubseteq_w),p\right\rangle\rightarrow\left\langle v,sto',(w',\sqsubseteq_w'),(L,V),p''\right\rangle$,
		\item $\Gamma,\Pi\vdash e^{p'} : T$
		\item $\Gamma,\Pi\vdash env$
		\item $(env,sto,(w,\sqsubseteq_w))\models(\Gamma,\Pi)$
	\end{itemize}
	Then we have that:
	\begin{itemize}
		\item $\Gamma,\Pi\vdash v:T$
		\item $(env,sto',(w',\sqsubseteq_w'))\models(\Gamma,\Pi)$
		\item $(env,(w',\sqsubseteq_w'),v,(L,V))\models(\Gamma,T)$
	\end{itemize}
\end{theorem}
\begin{proof}
	The proof proceeds by induction on the height of the derivation tree for 
	$$env\vdash\left\langle e^{p'},sto,\psi,p\right\rangle\rightarrow\left\langle v,sto',\psi',(L,V),p''\right\rangle$$
	We will only show the proof of four rules here, for \runa{Var}, \runa{Case}, \runa{Ref}, and \runa{Ref-write}, the full proof can be found in \cref{app:SoundnessProof}.

	\begin{description}
		\item[\runa{Var}] Here $e^{p'}=x^{p'}$, where
\begin{figure}[H]
	\setlength\tabcolsep{8pt}
	\begin{tabular}{l}
		\input{sections/appendix/ColRules/var.tex}
	\end{tabular}
\end{figure}
And from our assumptions, we have:
\begin{itemize}
	\item $\Gamma,\Pi\vdash x^{p'} : T$
	\item $\Gamma,\Pi\vdash env$
	\item $(env,sto,(w,\sqsubseteq_w))\models(\Gamma,\Pi)$
\end{itemize}
To type the occurrence $x^{p'}$ we use the rule \runa{T-Var}:
\begin{figure}[H]
	\setlength\tabcolsep{8pt}
	\begin{tabular}{l}
		\runa{T-Var}\\[0.2cm]
			\inference[]{}
			{\Gamma,\Pi \vdash x^p:T \sqcup (\{x^p\},\emptyset)}
	\end{tabular}
\end{figure}
Where $x^{p''}=uf_{\sqsubseteq_\Pi}(x,\Gamma)$, $\Gamma(x^{p''})=T$.

We need to show that \cat{1)} $\Gamma,\Pi\vdash c:T$, \cat{2)} $(env,sto',(w',\sqsubseteq_w'))\models(\Gamma,\Pi)$, and\\
\cat{3)} $(env,v,(w',\sqsubseteq_w'),(L,V))\models(\Gamma,T)$.
\begin{description}
	\item[1)] Since, from our assumption, we know that $\Gamma,\Pi\vdash env$, we can then conclude that $\Gamma,\Pi\vdash v:T$

	\item[2)] Since there are no updates to $sto$ and $(w,\sqsubseteq_w)$, we then know from our assumptions that $(env,sto,(w,\sqsubseteq_w))\models(\Gamma,\Pi)$ holds after an evaluation.

	\item[3)] Since there are no updates to $sto'$ and $(w',\sqsubseteq_w')$, that $(L,V)$ is a result from looking up $x^{p''}$ in $(w,\sqsubseteq_w)$, and the type $T$ is from to looking up $x^{p''}$ in $\Gamma$, 
		we then know that $(env,v,(w',\sqsubseteq_w'),(L,V))\models(\Gamma,T)$.
		Due to \cref{def:TAgree} we can conclude that: 
		$$(env,v,(w',\sqsubseteq_w'),(L,V\cup\{x^{p''}\}))\models(\Gamma,T\sqcup \{x^{p''}\})$$
\end{description}

		\item[\runa{Case}] Here $e^{p'}=\left[\mbox{case}\;e^{p''}\;\tilde{\pi}\;\tilde{o}\right]^{p'}$, where
\begin{figure}[H]
	\setlength\tabcolsep{8pt}
	\begin{tabular}{l}
		\input{sections/appendix/ColRules/case.tex}
	\end{tabular}
\end{figure}

And from our assumptions, we have that:
\begin{itemize}
	\item $\Gamma,\Pi\vdash \left[\mbox{case}\;e^{p''}\;\tilde{\pi}\;\tilde{o}\right]^{p'}:T$,
	\item $\Gamma,\Pi\vdash env$
	\item $(env,sto,(w,\sqsubseteq_w))\models(\Gamma,\Pi)$,
\end{itemize}
To type $[\mbox{case}\;e^{p''}\;\tilde{\pi}\;\tilde{o}]^{p'}$ we need to use the \runa{T-Case} rule, where we have:
\begin{figure}[H]
	\setlength\tabcolsep{8pt}
	\begin{tabular}{l}
		\runa{T-Case}\\[0.2cm]
			\inference[]
				{\Gamma,\Pi\vdash e^{p}:(\delta,\kappa) &\\
				\Gamma',\Pi\vdash e_i^{p_i}:T_i\;\;\;(1\leq i\leq|\tilde{\pi}|)}
				{\Gamma,\Pi\vdash [\mbox{case}\;e^{p}\;\tilde{\pi}\;\tilde{o}]^{p'}:T}
	\end{tabular}
\end{figure}
Where $T=T'\sqcup(\delta,\kappa)$, $T'=\bigcup_{i=1}^{|\tilde{\pi}|}T_i$, $e_i^{p_i}\in\tilde{o}$ and $s_i\in\tilde{\pi}$, and $\Gamma'=\Gamma[x^p:(\delta,\kappa)]$ if $s_i=x$.

We must show that \cat{1)} $\Gamma,\Pi\vdash v:T$, \cat{2)} $(env,sto',(w',\sqsubseteq_w'))\models(\Gamma,\Pi)$, and \\
\cat{3)} $(env,v,(w',\sqsubseteq_w'),(L,V))\models(\Gamma,T)$.

To conclude, we first need to show for the premises, where due to our assumption and from the first premise, we can use the induction hypothesis to get:
\begin{itemize}
	\item $\Gamma,\Pi\vdash v_e:(\delta,\kappa)$,
	\item $(env,sto'',(w'',\sqsubseteq_w''))\models(\Gamma,\Pi)$,
	\item $(env,v,(w'',\sqsubseteq_w''),(L,V))\models(\Gamma,(\delta,\kappa))$
\end{itemize}
Since in the rule \runa{T-Case} we take the union of all patterns, we can then from the second premise:
\begin{itemize}
	\item $\Gamma,\Pi\vdash v:T_j$,
	\item $(env,sto',(w',\sqsubseteq_w'))\models(\Gamma,\Pi)$,
	\item $(env,v,(w',\sqsubseteq_w'),(L,V))\models(\Gamma,T_j)$
\end{itemize}

If we have \cat{a)} $\Gamma',\Pi\vdash env[env']$ and \cat{b)} $(env[env'],sto'',(w''',\sqsubset_w''))\models(\Gamma',\Pi)$, we can then conclude the second premise by our induction hypothesis.
\begin{description}
	\item[a)] We know that either we have $\Gamma'=\Gamma[x\mapsto(\delta,\kappa)]$ and $env[x\mapsto v_e]$ if $s_j=x$, or $\Gamma'=\Gamma$ and $env$ if $s_j\neq x$.
		\begin{itemize}
			\item if $s_j\neq x$: Then we have $\Gamma,\Pi\vdash env$
			\item if $s_j=x$: Then we have $\Gamma[x\mapsto(\delta,\kappa)],\Pi\vdash env[x\mapsto v_e]$, which hold due to the first premise.
		\end{itemize}
	\item[b)] We know that either we have $\Gamma'=\Gamma[x\mapsto(\delta,\kappa)]$ and $env[x\mapsto v_e]$ if $s_j=x$, or $\Gamma'=\Gamma$ and $env$ if $s_j\neq x$.
		\begin{itemize}
			\item if $s_j\neq x$: then we have $(env,sto'',(w'',\sqsubset_w''))\models(\Gamma,\Pi)$.
			\item if $s_j=x$: then $(env[x\mapsto v_e],sto'',(w''',\sqsubset_w''))\models(\Gamma[x\mapsto(\delta,\kappa)],\Pi)$, since we know that $(env,sto'',(w'',\sqsubset_w''))\models(\Gamma,\Pi)$, we only need to show for $x$.
				Since we have $x\in dom(env)$, $x^{p_j}\in dom(w''')$ and $x^{p_j}\in dom(\Gamma')$ and due to the first premise, we know that $(env[x\mapsto v_e],sto'',(w''',\sqsubset_w''))\models(\Gamma[x\mapsto(\delta,\kappa)],\Pi)$.
		\end{itemize}
\end{description}
Based on \cat{a)} and \cat{b)} we can then conclude:

\begin{description}
	\item[1)] Since $\Gamma',\Pi\vdash v:T_j$, then we also must have $\Gamma',\Pi\vdash v:T$, since $T$ only contains more information than $T_j$.
	\item[2)] By the second premise, \cref{lemma:His}, and \cref{lemma:Strength}, we can then get 
		$$(env,sto',(w',\sqsubseteq_w'))\models(\Gamma,\Pi)$$
	\item[3)] Due to \cat{1)}, \cat{2)}, \cat{a)}, and \cat{b)} we can then conclude that
		$$(env,v,(w',\sqsubseteq_w'),(L,V))\models(\Gamma,T)$$
\end{description}

		\item[\runa{Ref-read}] Here $e^{p'}=[!e_1^{p_1}]^{p'}$, where
\begin{figure}[H]
	\setlength\tabcolsep{8pt}
	\begin{tabular}{l}
		\input{sections/appendix/ColRules/refread.tex}
	\end{tabular}
\end{figure}
And from our assumptions, we have that:
\begin{itemize}
	\item $\Gamma,\Pi\vdash [!e_1^{p_1}]^{p'}:T$,
	\item $\Gamma;\Pi\vdash env$
	\item $(env,sto,(w,\sqsubseteq_w))\models(\Gamma,\Pi)$,
\end{itemize}
To type $[!e_1^{p_1}]^{p'}$ we need to use the \runa{T-Ref-read} rule, where we have:
\begin{figure}[H]
	\setlength\tabcolsep{8pt}
	\begin{tabular}{l}
		\runa{T-Ref-read}\\[0.2cm]
			\inference[]
				{\Gamma,\Pi\vdash  e^{p}:(\delta,\kappa)}
				{\Gamma,\Pi\vdash [!e^{p}]^{p'}:T\sqcup(\delta\cup\delta',\emptyset)}\\
	\end{tabular}
\end{figure}
Where $\kappa\neq\emptyset$, $\delta'=\{\nu x^{p'}\mid\nu x\in\kappa\}$, $\nu x_1,\cdots,\nu x_n\in\kappa$.\\ 
$\{\nu x_1^{p_1},\cdots,\nu x_1^{p_m}\}=uf_{\Upsilon_{p'}}(\nu x_1,\Gamma),\cdots,\{\nu x_n^{p_1'},\cdots,\nu x_n^{p_s'}\}=uf_{\Upsilon_{p'}}(\nu x_n,\Gamma)$, and\\
$T=\Gamma(\nu x_1^{p_1})\cup\cdots\cup\Gamma(\nu x_1^{p_m})\cup\cdots\cup\Gamma(\nu x_n^{p_1'})\cup\cdots\cup\Gamma(\nu x_n^{p_s'})$.

We must show that \cat{(1)} $\Gamma,\Pi\vdash v:T$, \cat{(2)} $(env,sto',(w',\sqsubseteq_w'))\models(\Gamma,\Pi)$, and\\
\cat{(3)} $(env,v,(w',\sqsubseteq_w'),(L,V))\models(\Gamma,T)$.

To conclude, we first need to show for the premises, where due to our assumption and from the premise, we can use the induction hypothesis to get:
\begin{itemize}
	\item $\Gamma,\Pi\vdash \loc:(\delta,\kappa)$,
	\item $(env,sto',(w',\sqsubseteq_w'))\models(\Gamma,\Pi)$,
	\item $(env,v,(w',\sqsubseteq_w'),(L,V))\models(\Gamma,(\delta',\kappa'))$
\end{itemize}

Due to $(env,sto',(w',\sqsubseteq_w'))\models(\Gamma,\Pi)$ and $(env,v,(w',\sqsubseteq_w'),(L,V))\models(\Gamma,(\delta',\kappa'))$, and due to our assumptions, we can conclude that:
\begin{description}
	\item[(1)] $\Gamma,\Pi\vdash v:T$,

	\item[(2)] $(env,sto',(w',\sqsubseteq_w'))\models(\Gamma,\Pi)$,

	\item[(3)] $(env,v,(w',\sqsubseteq_w'),(L\cup\{\loc^{p''}\},V))\models(\Gamma,T\sqcup(\delta\cup\delta',\emptyset))$
\end{description}

	\end{description}
\end{proof}

%\subfile{SoundProof/index.tex}

\section{Conclusion}\label{sec:Conc}
In this paper, we have introduced a type system for local data-flow analysis for a language based on a subset of ReScript.
The type system we present here differs from other data-flow analysis techniques, that instead of solving constraints, gives a semantic analysis of a program.

In the type system, we have shown how to handle data-flow analysis for different language constructs, for pattern matching, local declarations, and referencing.
As pattern matching introduces branches to the language, we showed a sound over-approximation of how to handle these branches.
Additionally, since we also have mutability, through referencing, the approximation should also, in case of reading from a reference, get all places where a reference binding could exist in the type environment.
Since some branches could write to a reference, while others do not, it was important to consider each branch separately when reading from references.

\subsection{discussion}
The type system we have presented is for a small language without many constructs.
However, some interesting constructs were introduced to the language, such as mutability and pattern matching which introduces some challenges when trying to make a good approximation of data-flow.

The challenges from having both pattern matching and mutability introduced challenges, as each branch could not simply be thought of as locally, since reference operations introduce side effects.
To references, we represent them as internal variables, i.e., variables that do not exist in the syntax, and treating them as global information.
To handle the problem of branching, when reading from a reference, we look at each branch independently to find the information necessary.
\bigskip

Since we focused on a functional language, that is based on expressions, we focused the data-flow analysis on the flow of variables, i.e., which variables are used to evaluate an occurrence.
As the language is primarily a series of declaration of functions, variables, and references, this allows for analyzing where variables are used on which are useful evaluating an occurrence.

Additionally by representing referencing as internal variables, it allows for understanding of which references are used and where they are used in the occurrence.
This information can be used by compilers to make sure that references can be safely cleaned up, after the last place they were used.
The alias information also implies which aliases were used in the occurrence.
\bigskip

However, the type system introduced contains some restrictions, also called slack, for which occurrences it accepts.
As the type system does not allow for type polymorphism, the use cases for abstractions are restricted.
In one place, where abstractions are quite limited is when binding them to a local declaration, this local declaration cannot be used at multiple places, as this would mean it would contain occurrences at multiple program points.

Another area of the type system contains slack, is for references as abstractions cannot be bound to them.
Here, another issue occurs as the environment only contains local information, and an abstraction thus only knows about the variables known when it was declared.
As the type system is currently defined, the type environment should be bound with the abstraction, but the current type system does not allow this, as the type environment both includes local and global information.

\subsection{Future work}\label{sec:FW}
We will now introduce potential future work, as areas of improvement for the type system

\paragraph{Implementation of a type checker}
Implementing a type checker for the type system presented here would allow for testing how well the information is used.
It would also allow for comparing how well it performs, compared to other data-flow analysis techniques.

\paragraph{Polymorphism}
Introducing polymorphism would be an ideal place to extend upon the type system, as abstractions are restricted in the current type system.
Here, polymorphism for the base type, that is for $(\delta,\kappa)$, would allow for abstractions to be used multiple times in an occurrence, the input and output type would not be restricted from only allowing the exact same input type.
As such, consider the following occurrence:
\begin{lstlisting}[language=Caml, mathescape=true]
(let x ($\lambda$ y.y$^1$)$^2$ (x$^3$ (x$^4$ 1$^5$)$^6$)$^7$)$^8$
\end{lstlisting}
If polymorphism is introduced, occurrences like this could be defined, since when typing the applications, the type of the argument changes, since the occurrence $x^4$ is present in the second application.

\paragraph{Extending references}
References are defined currently in the type system, they cannot be bound to abstractions.
However, this would also introduce complications, as abstractions need the information known at the time they were declared.
Another complication would be that if different references had different types, e.g., if it had an abstraction type at one point and a base type at another point.
Here, either we should require references to always have the same type, e.g., with base type polymorphism.
Consider the following occurrence:
\begin{lstlisting}[language=Caml, mathescape=true]
((!(case 1 (1$^1$) (let z 5$^2$ (ref ($\lambda$ y.(PLUS z$^3$ y$^4$)$^5$)$^6$))$^7$)$^8$)$^9$ 5)$^{10}$
\end{lstlisting}
This occurrence would create a reference to a local abstraction which depends on the locally declared variable $z$ before reading from the reference and applying the constant to it.
In the semantics, the environment would be added to the abstraction closure, and when evaluating the body of the abstraction, in an application, it would use the environment in the closure.

\paragraph{Type inference}
Another area is to make a type inference algorithm, which can find the type information.
To make type inference for the type system would need to find an approximated order of program points, find a proper $\kappa_0$ and type for abstractions, that is, find all the places where the parameter should be bound.

\paragraph{Extending with more language constructs}
It would also be interesting to introduce more language constructs, as the language presented only contains a small amount of constructs, such as mutability and pattern matching.
Some interesting constructs to add could be records, constructors and deconstructors, modules, or lazy evaluation.
Here, lazy evaluation could take multiple forms, either by introducing it as a core part of the language, where every binding is lazy evaluated, or add special constructs for lazy evaluation.
Modules, on the other hand, would allow for wrapping an occurrence, or multiple occurrences into a module, which could then be used in multiple places.

	\balance
	\bibliographystyle{ACM-Reference-Format}
	\bibliography{bib.bib}
	\appendix

\section{Collection Semantics}\label{App:ColSem}
\begin{figure}[H]
	\setlength\tabcolsep{8pt}
	\begin{tabular}{l}
		\input{sections/appendix/ColRules/const.tex}\\[1cm]
		\input{sections/appendix/ColRules/var.tex}\\[1cm]
		\input{sections/appendix/ColRules/abs.tex}\\[1cm]
		\input{sections/appendix/ColRules/app.tex}\\[1cm]
		\input{sections/appendix/ColRules/apprec.tex}\\[1cm]
		\input{sections/appendix/ColRules/appconst.tex}
	\end{tabular}
	\label{fig:InfDV}
\end{figure}

\begin{figure}[H]
	\setlength\tabcolsep{8pt}
	\begin{tabular}{l}
		\input{sections/appendix/ColRules/let.tex}\\[1cm]
		\input{sections/appendix/ColRules/letrec.tex}\\[1cm]
		\input{sections/appendix/ColRules/case.tex}\\[1cm]
		\input{sections/appendix/ColRules/ref.tex}\\[1cm]
		\input{sections/appendix/ColRules/refread.tex}\\[1cm]
		\input{sections/appendix/ColRules/refwrite.tex}
	\end{tabular}
	\label{fig:InfDV}
\end{figure}

\subsection{Pattern matching}
Pattern matching matches the first expression, $e$, with each pattern to find a match.
Here, we define the function $match:Val \times Pat \rightarrow (\Var \rightharpoonup \Exp)$, where $Pat$ is the set of patterns.

The $match(v,s)=env$ function, where we get a substitution $env$.
We defined the function inductively as follows:

\begin{align*}
	match(n,n) &= id\\
	match(b,b) &= id\\
	match(v,\_) &=id\\
	match(e,x) &= [x \mapsto e]\\
	match(\_,p) &= \perp
\end{align*}

\paragraph{number and boolean}
Matching of numbers and booleans are an equality match, so those cases returns the identity.

\paragraph{variables}
Variable pattern matching instantiates the pattern, by binding the expression to the variable.

\paragraph{patterns}
Matching a record matches on all patterns in the expression $e$, where $I$ denote a finite amount of records fields.
Since there are multiple record fields, each of those need to be instantiated
$\sigma_i=match(\{l_i=e_i\}^n_{i \in I},\{l_i\}^n_{i \in I})$

\paragraph{wildcard and fail}
The last two patterns to match are the wildcard and fail cases.
The wildcard accepts any input $e$ and gives an emtpy sybstitution, the fail case just return a false boolean value, since it were an unsuccessful match.

\subsection{Extending $w$}
Similarly to the pattern matching function, we define a function that extends the dependency function for binding pattern bindings to it liveness information.
This is similarly defined, where $match_w(s,p,(L,V))=\psi$ is a function that returns an extension $\psi$.
The $match_w$ function can thus be defined by:
\begin{align}
	match_w(s,p,(L,V))) =
	\left\{\begin{matrix}
		[x^p\mapsto (L,V)] & \mbox{if}\; s=x\\ 
		[] & \mbox{otherwise}
	\end{matrix}\right.
\end{align}

\section{Type system Judgement}\label{App:TypeSys}
\begin{figure}[H]
	\setlength\tabcolsep{8pt}
	\begin{tabular}{l}
		\input{sections/appendix/TSRules/const.tex}\\[1cm]
		\input{sections/appendix/TSRules/var.tex}\\[1cm]
		\input{sections/appendix/TSRules/abs.tex}\\[1cm]
		\input{sections/appendix/TSRules/app.tex}\\[1cm]
		\input{sections/appendix/TSRules/appconst.tex}\\[1cm]
		\input{sections/appendix/TSRules/let1.tex}\\[1cm]
		\input{sections/appendix/TSRules/let2.tex}\\[1cm]
	\end{tabular}
	\label{fig:TypeSys1}
\end{figure}

\begin{figure}[H]
	\setlength\tabcolsep{8pt}
	\begin{tabular}{l}
		\input{sections/appendix/TSRules/letrec.tex}\\[1cm]
		\input{sections/appendix/TSRules/case.tex}\\[1cm]
		\input{sections/appendix/TSRules/ref.tex}\\[1cm]
		\input{sections/appendix/TSRules/refread.tex}\\[1cm]
		\input{sections/appendix/TSRules/refwrite.tex}
	\end{tabular}
	\label{fig:TypeSys2}
\end{figure}

\subsection{$\sigma$ function}
The $\sigma$ function is used to extend $\Gamma$ in case of a variable pattern, which takes as input $\Gamma$, a pattern $s$, a program point $p$, and a type $T$.
$\sigma$ can then inductively be defined as:
\begin{align*}
	\sigma(n,T) &= id\\
	\sigma(b,T) &= id\\
	\sigma(\_,T) &= id\\
	\sigma(x,T) &= [x^p:T]
\end{align*}

\section{Proofs of theorems and lemmas}
\subsection{History}\label{app:HisProof}
Here, we present the proof for the history, that is, all variables introduced in an evaluation of an occurrence, and bound to the dependency function, is not free.

\begin{lemma}[History]
	Suppose $e^p$ is an occurrence, that
	$$env\vdash\left\langle e^{p},sto,(w,\sqsubseteq_w),p'\right\rangle\rightarrow\left\langle v,sto',(w',\sqsubseteq_w'),(L,V),p''\right\rangle$$
		and $x^{p_1}\in dom(w')\backslash dom(w)$.
		Then $x\notin fv(e^{p})$
\end{lemma}

\begin{proof}
	The proof proceeds by induction on the height of the derivation tree for 
	$$env\vdash\left\langle e^{p'},sto,\psi,p\right\rangle\rightarrow\left\langle v,sto',\psi',(L,V),p''\right\rangle$$
	In the base case, we have \runa{Const}, \runa{Var}, and \runa{Abs}:
	\begin{description}
		\item[\runa{Cons}] Here $e^{p'}=c^{p'}$, where
\begin{figure}[H]
	\setlength\tabcolsep{8pt}
	\begin{tabular}{l}
		\runa{Const}\\[0.2cm]
			\inference[]{}
			{env\vdash \left\langle c^{p'},sto,(w,\sqsubseteq_w),p \right\rangle \rightarrow \left\langle c,sto,(w,\sqsubseteq_w),(\emptyset,\emptyset),p' \right\rangle}
	\end{tabular}
\end{figure}
Since there is no updates to $w$, this case follows immediately.

		\item[\runa{Var}] Here $e^{p'}=x^{p'}$, where

\begin{figure}[H]
	\setlength\tabcolsep{8pt}
	\begin{tabular}{l}
		\input{sections/appendix/ColRules/var.tex}
	\end{tabular}
\end{figure}
Since there are no updates to $w$, this case follows immediately.

		\item[\runa{Abs}] Here $e^{p'}=[\lambda\;x.e^{p''}]^{p'}$, where
\begin{figure}[H]
	\setlength\tabcolsep{8pt}
	\begin{tabular}{l}
		\input{sections/appendix/ColRules/abs.tex}
	\end{tabular}
\end{figure}
Since there is no updates to $w$, this case follows immediately.

	\end{description}

	Next, follows the induction step:
	\begin{description}
		\item[\runa{App}] Here $e^{p'}=[e_1^{p_1}\;e_2^{p_2}]^{p'}$, where
\begin{figure}[H]
	\setlength\tabcolsep{8pt}
	\begin{tabular}{l}
		\input{sections/appendix/ColRules/app.tex}
	\end{tabular}
\end{figure}
By virtue of our induction hypothesis, we can get the follow from the premises:
\begin{description}
	\item[1)] if $y^{p''}\in dom(w_1)\backslash dom(w)$ then $y\notin fv(e_1^{p_1})$
	\item[2)] if $y^{p''}\in dom(w_2)\backslash dom(w_1)$ then $y\notin fv(e_2^{p_2})$
	\item[3)] if $y^{p''}\in dom(w')\backslash dom(w_3)$ then $y\notin fv(e_3^{p_3})$
\end{description}
We the need to show that: if $y^{p''}\in dom(w')\backslash dom(w)$ then $y\notin fv(e^{p'})$.
By \cref{def:fv} we know that $fv([e_1^{p_1}\;e_2^{p_2}]^{p'})=fv(e_1^{p_1})\cup fv(e_2^{p_2})$, and the only variable that is not handled by \cat{1)}, \cat{2)}, and \cat{3)} is $x^{p_2}$.
But since $x$ is not a free variable in $e_1^{p_1}$ or $e_2^{p_2}$, this case then follows.

		\item[\runa{App-const}] Here $e^{p'}=[e_1^{p_1}\;e_2^{p_2}]^{p'}$, where
\begin{figure}[H]
	\setlength\tabcolsep{8pt}
	\begin{tabular}{l}
		\input{sections/appendix/ColRules/appconst.tex}
	\end{tabular}
\end{figure}
By virtue of our induction hypothesis, we can get the following from the premises:
\begin{description}
	\item[1)] if $y^{p''}\in dom(w_1)\backslash dom(w)$ then $y\notin fv(e_1^{p_1})$
	\item[2)] if $y^{p''}\in dom(w')\backslash dom(w_1)$ then $y\notin fv(e_2^{p_2})$
\end{description}
We the need to show that: if $y^{p''}\in dom(w')\backslash dom(w)$ then $y\notin fv(e^{p'})$.
By \cref{def:fv} we know that $fv([e_1^{p_1}\;e_2^{p_2}]^{p'})=fv(e_1^{p_1})\cup fv(e_2^{p_2})$, this case then follows.

		\item[\runa{App-rec}] Here $e^{p'}=[e_1^{p_1}\;e_2^{p_2}]^{p'}$, where
\begin{figure}[H]
	\setlength\tabcolsep{8pt}
	\begin{tabular}{l}
		\input{sections/appendix/ColRules/apprec.tex}
	\end{tabular}
\end{figure}
By virtue of our induction hypothesis, we can get the following from the premises:
\begin{description}
	\item[1)] if $y^{p''}\in dom(w_1)\backslash dom(w)$ then $y\notin fv(e_1^{p_1})$
	\item[2)] if $y^{p''}\in dom(w_2)\backslash dom(w_1)$ then $y\notin fv(e_2^{p_2})$
	\item[3)] if $y^{p''}\in dom(w')\backslash dom(w_3)$ then $y\notin fv(e_3^{p_3})$
\end{description}
We the need to show that: if $y^{p''}\in dom(w')\backslash dom(w)$ then $y\notin fv(e^{p'})$.
By \cref{def:fv} we know that $fv([e_1^{p_1}\;e_2^{p_2}]^{p'})=fv(e_1^{p_1})\cup fv(e_2^{p_2})$, and the only variable that is not handled by \cat{1)}, \cat{2)}, and \cat{3)} is $x^{p_2}$.
But since $x$ is not a free variable in $e_1^{p_1}$ or $e_2^{p_2}$, this case follows.

		\item[\runa{Let}] Here $e^{p'}=[\mbox{let}\;x\;e_1^{p_1}\;e_2^{p_2}]^{p'}$, where
\begin{figure}[H]
	\setlength\tabcolsep{8pt}
	\begin{tabular}{l}
		\input{sections/appendix/ColRules/let.tex}
	\end{tabular}
\end{figure}
By virtue of our induction hypothesis, we can get the following from the premises:
\begin{description}
	\item[1)] if $y^{p''}\in dom(w_1)\backslash dom(w)$ then $y\notin fv(e_1^{p_1})$
	\item[2)] if $y^{p''}\in dom(w')\backslash dom(w_1)$ then $y\notin fv(e_2^{p_2})$
\end{description}
We then need to show that: if $y^{p''}\in dom(w')\backslash dom(w)$ then $y\notin fv(e^{p'})$.
By \cref{def:fv} we know that $fv([\mbox{let}\;x\;e_1^{p_1}\;e_2^{p_2}]^{p'})=fv(e_1^{p_1})\cup fv(e_2^{p_2})\backslash\{x\}$, and the only variable that is not handled by \cat{1)}, \cat{2)}, and \cat{3)} is $x^{p_2}$.
Where $x$ is not a free variable in $e_1^{p_1}$, but $x$ is possibly free in $e_2^{p_2}$.
From \cref{def:fv}, we know that $x$ is not free in $[\mbox{let}\;x\;e_1^{p_1}\;e_2^{p_2}]^{p'}$, we then get: if $y^{p''}\in dom(w')\backslash dom(w)$ then $y\notin fv(e^{p'})$.

		\item[\runa{Let-rec}] Here $e^{p'}=[\mbox{let rec}\;x\;e_1^{p_1}\;e_2^{p_2}]^{p'}$, where
\begin{figure}[H]
	\setlength\tabcolsep{8pt}
	\begin{tabular}{l}
		\input{sections/appendix/ColRules/letrec.tex}
	\end{tabular}
\end{figure}
By virtue of our induction hypothesis, we can get the following from the premises:
\begin{description}
	\item[1)] if $y^{p''}\in dom(w_1)\backslash dom(w)$ then $y\notin fv(e_1^{p_1})$
	\item[2)] if $y^{p''}\in dom(w')\backslash dom(w_1)$ then $y\notin fv(e_2^{p_2})$
\end{description}
We then need to show that: if $y^{p''}\in dom(w')\backslash dom(w)$ then $y\notin fv(e^{p'})$.
By \cref{def:fv} we know that $fv([\mbox{let rec}\;f\;e_1^{p_1}\;e_2^{p_2}]^{p'})=fv(e_1^{p_1})\cup fv(e_2^{p_2})\backslash\{f\}$, and the only variable that is not handled by \cat{1)}, \cat{2)}, and \cat{3)} is $f^{p_2}$.
Where $f$ is possible free in $e_1^{p_1}$ and $e_2^{p_2}$.
From \cref{def:fv}, we know that $f$ is not free in $[\mbox{let rec}\;f\;e_1^{p_1}\;e_2^{p_2}]^{p'}$, we then get: if $y^{p''}\in dom(w')\backslash dom(w)$ then $y\notin fv(e^{p'})$.

		\item[\runa{Case}] Here $e^{p'}=[\mbox{let}\;x\;e_1^{p_1}\;e_2^{p_2}]^{p'}$, where
\begin{figure}[H]
	\setlength\tabcolsep{8pt}
	\begin{tabular}{l}
		\input{sections/appendix/ColRules/case.tex}
	\end{tabular}
\end{figure}
By virtue of our induction hypothesis, we can get the following from the premises:
\begin{description}
	\item[1)] if $y^{p''}\in dom(w'')\backslash dom(w)$ then $y\notin fv(e^{p''})$
	\item[2)] if $y^{p''}\in dom(w')\backslash dom(w''')$ then $y\notin fv(e_j^{p_j})$
\end{description}
We then need to show that: if $y^{p''}\in dom(w')\backslash dom(w)$ then $y\notin fv(e^{p'})$.
By \cref{def:fv} we know that $fv([\mbox{case}\;e^{p''}\;\tilde{\pi}\;\tilde{o}]^{p'})=fv(e_1^{p_j})\cup\cdots\cup fv(e_n^{p_n})\backslash(\tau(s_1)\cup\cdots\cup\tau(s_n))$, this case then follows.

		\item[\runa{Ref}] Here $e^{p'}=[\mbox{let}\;x\;e_1^{p_1}\;e_2^{p_2}]^{p'}$, where
\begin{figure}[H]
	\setlength\tabcolsep{8pt}
	\begin{tabular}{l}
		\input{sections/appendix/ColRules/ref.tex}
	\end{tabular}
\end{figure}
By virtue of our induction hypothesis, we can get the following from the premises:
\begin{description}
	\item[1)] if $y^{p''}\in dom(w')\backslash dom(w)$ then $y\notin fv(e_1^{p_1})$
\end{description}
By \cref{def:fv} we can then conclude that: if $y^{p''}\in dom(w')\backslash dom(w)$ then $y\notin fv(e^{p'})$

		\item[\runa{Ref-read}] Here $e^{p'}=[\mbox{let}\;x\;e_1^{p_1}\;e_2^{p_2}]^{p'}$, where
\begin{figure}[H]
	\setlength\tabcolsep{8pt}
	\begin{tabular}{l}
		\input{sections/appendix/ColRules/refread.tex}
	\end{tabular}
\end{figure}
By virtue of our induction hypothesis, we can get the following from the premises:
\begin{description}
	\item[1)] if $y^{p''}\in dom(w')\backslash dom(w)$ then $y\notin fv(e_1^{p_1})$
\end{description}
By \cref{def:fv} we can then conclude that: if $y^{p''}\in dom(w')\backslash dom(w)$ then $y\notin fv(e^{p'})$

		\item[\runa{Ref-write}] Here $e^{p'}=[\mbox{let}\;x\;e_1^{p_1}\;e_2^{p_2}]^{p'}$, where
\begin{figure}[H]
	\setlength\tabcolsep{8pt}
	\begin{tabular}{l}
		\input{sections/appendix/ColRules/refwrite.tex}
	\end{tabular}
\end{figure}
By virtue of our induction hypothesis, we can get the following from the premises:
\begin{description}
	\item[1)] if $y^{p''}\in dom(w_1)\backslash dom(w)$ then $y\notin fv(e_1^{p_1})$
	\item[2)] if $y^{p''}\in dom(w')\backslash dom(w_1)$ then $y\notin fv(e_2^{p_2})$
\end{description}
We then need to show that: if $y^{p''}\in dom(w')\backslash dom(w)$ then $y\notin fv(e^{p'})$.
By \cref{def:fv} we know that $fv([e_1^{p_1}:=e_2^{p_2}]^{p'})=fv(e_1^{p_1})\cup fv(e_2^{p_2})$.
From \cref{def:fv}, we then get: if $y^{p''}\in dom(w')\backslash dom(w)$ then $y\notin fv(e^{p'})$.

	\end{description}
\end{proof}

\subsection{Strengthening}\label{app:StrProof}
Here, we present the proof for strengthening of type environments.

\begin{lemma}[Strengthening]
	If $\Gamma[x^{p'}:T'],\Pi\vdash e^{p}:T$ and $x\notin fv(e^p)$, then $\Gamma,\Pi\vdash e^{p}:T$
\end{lemma}

\begin{proof}
	The proof proceeds by induction on the structure of the derivation tree for the type judgment:
	$$\Gamma,\Pi\vdash e^{p}:T$$
	In the base case, we have \runa{T-Const} and \runa{T-Var}:
	\begin{description}
		\item[\runa{T-Const}] Here $e^p=c^p$, where
\begin{figure}[H]
	\setlength\tabcolsep{8pt}
	\begin{tabular}{l}
		\input{sections/appendix/TSRules/const.tex}
	\end{tabular}
\end{figure}
and from our assumption, we know that $x\notin fv(c^p)$.
We can thus conclude that:
$$\Gamma;\Pi\vdash c^p : (\emptyset,\emptyset)$$

		\item[\runa{T-Var}] Here $e^p=x^p$, where
\begin{figure}[H]
	\setlength\tabcolsep{8pt}
	\begin{tabular}{l}
		\input{sections/appendix/TSRules/var.tex}
	\end{tabular}
\end{figure}
From our assumption, we know that $x\notin fv(x^p)$.
We can thus conclude that:
$$\Gamma,\Pi\vdash y^p : (\emptyset,\emptyset)$$

	\end{description}

	Next, follows the induction step:
	\begin{description}
		\item[\runa{T-Abs}] Here $e^p=[\lambda\;y.e_1^{p_1}]^p$, where
\begin{figure}[H]
	\setlength\tabcolsep{8pt}
	\begin{tabular}{l}
		\runa{T-Abs}\\[0.2cm]
			\inference[]
				{\Gamma'[y^{p_0}:T_1],\Pi\vdash  e_1^{p_1}:T_2}
				{\Gamma',\Pi\vdash  [\lambda\;y.e_1^{p_1}]^{p}:T_1\rightarrow T_2}\\[0.3cm]
	\end{tabular}
\end{figure}
Where $\Gamma'=\Gamma[x^{p'}:T_1]$ and $p\sqsubseteq_\Pi p_0\wedge p_0\sqsubseteq_\Pi p_1$.
From our assumption, we know that $x\notin fv(e^p)$.
By \cref{def:fv} we know that $x\notin fv(e_1^{p_1})$ where from our induction hypothesis we get that:
$$\Gamma[y^{p_0}:T_1],\Pi\vdash  e_1^{p_1}:T_2$$

We can thus conclude that:
$$\Gamma,\Pi\vdash[\lambda\;y.e_1^{p_1}]^{p}:T_1\rightarrow T_2$$

		\item[\runa{T-App}] Here $e^p=[e_1^{p_1} \; e_2^{p_2}]^p$, where
\begin{figure}[H]
	\setlength\tabcolsep{8pt}
	\begin{tabular}{l}
		\runa{T-App}\\[0.2cm]
		\inference[]
		{
			\Gamma',\Pi\vdash e_1^{p}:T_1\rightarrow T_2 &\\
			\Gamma',\Pi\vdash e_2^{p'}:T_1
		}
		{\Gamma',\Pi\vdash [e_1^{p} \; e_2^{p'}]^{p''}:T_2}
	\end{tabular}
\end{figure}
Where $\Gamma'=\Gamma[x^{p'}:T_1]$.
From our assumption, we know that $x\notin fv(e^p)$.
By \cref{def:fv}, we then know:
\begin{itemize}
	\item $x\notin fv(e_1^{p_1})$,
	\item $x\notin fv(e_2^{p_2})$
\end{itemize}
Then from our induction hypothesis we can get:
\begin{itemize}
	\item $\Gamma,\Pi\vdash e_1^{p_1}:T_1\rightarrow T$,
	\item $\Gamma,\Pi\vdash e_2^{p_2}:T_1$
\end{itemize}
Where we can then conclude that:
$$\Gamma,\Pi\vdash [e_1^{p_1} \; e_2^{p_2}]^{p}:T_2$$

		\item[\runa{T-App-const}] Here $e^p=[c\;e_1^{p_1} \; e_2^{p_2}]^p$, where
\begin{figure}[H]
	\setlength\tabcolsep{8pt}
	\begin{tabular}{l}
	\runa{T-App-const}\\[0.2cm]
	\inference[]
	{
		\Gamma,\Pi\vdash e_1^{p}:(\delta_1,\emptyset) &\\
		\Gamma,\Pi\vdash e_2^{p'}:(\delta_2,\emptyset)
	}
	{\Gamma,\Pi\vdash [c\;e_1^{p} \; e_2^{p'}]^{p''}:(\delta_1\cup\delta_2,\emptyset)}\\[0.3cm]
	\end{tabular}
\end{figure}
Where $\Gamma'=\Gamma[x^{p'}:T_1]$, and $c$ is a functional constant.
From our assumption, we know that $x\notin fv(e^p)$.
By \cref{def:fv}, we then know:
\begin{itemize}
	\item $x\notin fv(e_1^{p_1})$,
	\item $x\notin fv(e_2^{p_2})$
\end{itemize}
Then from our induction hypothesis we can get:
\begin{itemize}
	\item $\Gamma,\Pi\vdash e_1^{p_1}:(\delta_1,\emptyset)$,
	\item $\Gamma,\Pi\vdash e_2^{p_2}:(\delta_2,\emptyset)$
\end{itemize}
Where we can then conclude that:
$$\Gamma,\Pi\vdash [c\;e_1^{p_1} \; e_2^{p_2}]^{p}:T_2$$

		\item[\runa{T-Let-1}] Here $e^p=[\mbox{let}\; y \; e_1^{p_1} \; e_2^{p_2}]^p$, where
\begin{figure}[H]
	\setlength\tabcolsep{8pt}
	\begin{tabular}{l}
	\runa{T-Let-1}\\[0.2cm]
	\inference[]
	{
		\Gamma',\Pi\vdash e_1^{p_1}:(\delta,\kappa) &\\
		\Gamma',\Pi\vdash e_2^{p_2}:T_2
	}
	{\Gamma'',\Pi\vdash [\mbox{let}\; x \; e_1^{p_1} \; e_2^{p_2}]^{p}:T_2}\\[0.3cm]
	\end{tabular}
\end{figure}
Where $\Gamma'=\Gamma''[x^{p}:(\delta,\kappa\cup \{x\})]$ and $\kappa\neq\emptyset$, and $\Gamma''=\Gamma[x^{p'}:T_1]$.
From our assumption, we know that $x\notin fv(e^p)$.
By \cref{def:fv} we know:
\begin{itemize}
	\item $x\notin fv(e_1^{p_1})$,
	\item $x\notin fv(e_2^{p_2})$
\end{itemize}
By our induction hypothesis we can get:
\begin{itemize}
	\item $\Gamma,\Pi\vdash e_1^{p_1}:(\delta,\kappa)$,
	\item $\Gamma[y^{p_1}:(\delta,\kappa\cup \{y\})],\Pi\vdash e_2^{p_2}:T$
\end{itemize}
Where we can then conclude that:
$$\Gamma,\Pi\vdash [\mbox{let}\; y \; e_1^{p_1} \; e_2^{p_2}]^{p}:T$$

		\item[\runa{T-Let-2}] Here $e^p=[\mbox{let}\; y \; e_1^{p_1} \; e_2^{p_2}]^p$, where
\begin{figure}[H]
	\setlength\tabcolsep{8pt}
	\begin{tabular}{l}
	\runa{T-Let-2}\\[0.2cm]
	\inference[]
	{
		\Gamma',\Pi\vdash e_1^{p_1}:T_1 &\\
		\Gamma'[x^p:T_1],\Pi\vdash e_2^{p_2}:T_2
	}
	{\Gamma',\Pi\vdash [\mbox{let}\; x \; e_1^{p_1} \; e_2^{p_2}]^{p'}:T_2}
	\end{tabular}
\end{figure}
Where $\Gamma'=\Gamma[x^{p'}:T_1]$.
From our assumption, we know that $x\notin fv(e^p)$.
By \cref{def:fv} we know:
\begin{itemize}
	\item $x\notin fv(e_1^{p_1})$,
	\item $x\notin fv(e_2^{p_2})$
\end{itemize}
By our induction hypothesis we can get:
\begin{itemize}
	\item $\Gamma,\Pi\vdash e_1^{p_1}:T_1$,
	\item $\Gamma[y^{p_1}:T_1],\Pi\vdash e_2^{p_2}:T$
\end{itemize}
Where we can then conclude that:
$$\Gamma,\Pi\vdash [\mbox{let}\; y \; e_1^{p_1} \; e_2^{p_2}]^{p}:T$$

		\item[\runa{T-Let-rec}] Here $e^p=[\mbox{let rec}\; y \; e_1^{p_1} \; e_2^{p_2}]^p$, where
\begin{figure}[H]
	\setlength\tabcolsep{8pt}
	\begin{tabular}{l}
	\runa{T-Let-rec}\\[0.2cm]
	\inference[]
	{
		\Gamma',\Pi\vdash e_1^{p}:T\rightarrow T_2 &\\
		\Gamma'[f^p:T\rightarrow T_2],\Pi\vdash e_2^{p}:T_2
	}
	{\Gamma',\Pi\vdash [\mbox{let rec}\; f \; e_1^{p} \; e_2^{p'}]^{p''}:T_2}
	\end{tabular}
\end{figure}
Where $\Gamma'=\Gamma[x^{p'}:T_1]$.
From our assumption, we know that $x\notin fv(e^p)$.
By \cref{def:fv} we know:
\begin{itemize}
	\item $x\notin fv(e_1^{p_1})$,
	\item $x\notin fv(e_2^{p_2})$
\end{itemize}
By our induction hypothesis we can get:
\begin{itemize}
	\item $\Gamma,\Pi\vdash e_1^{p_1}:T_1\rightarrow T_2$,
	\item $\Gamma[f^{p_1}:T_1\rightarrow T_2],\Pi\vdash e_2^{p_2}:T$
\end{itemize}
Where we can then conclude that:
$$\Gamma,\Pi\vdash [\mbox{let rec}\; f \; e_1^{p_1} \; e_2^{p_2}]^{p}:T_2$$

		\item[\runa{T-Case}] Here $e^p=[\mbox{case}\;e^{p''}\;\tilde{\pi}\;\tilde{o}]^p$, where
\begin{figure}[H]
	\setlength\tabcolsep{8pt}
	\begin{tabular}{l}
	\runa{T-Case}\\[0.2cm]
	\inference[]
	{
		\Gamma',\Pi\vdash e^{p}:(\delta,\kappa) &\\
		\Gamma'',\Pi\vdash e_i^{p_i}:T_i\;\;\;(1\leq i\leq|\tilde{\pi}|)
	}
	{\Gamma',\Pi\vdash [\mbox{case}\;e^{p}\;\tilde{\pi}\;\tilde{o}]^{p'}:T\sqcup(\delta,\kappa)}\\[0.3cm]
	\end{tabular}
\end{figure}
Where $\Gamma'=\Gamma[x^{p'}:T_1]$, $e_i^{p_i}\in\tilde{o}$ and $s_i\in\tilde{\pi}$ $T=\bigcup_{i=1}^{|\tilde{\pi}|}T_i$, and
$\Gamma''=\Gamma'[x^p:(\delta,\kappa)]$ if $s_i=x$

From our assumption, we know that $x\notin fv(e^p)$.
By \cref{def:fv} we know:
\begin{itemize}
	\item $x\notin fv(e^{p''})$,
	\item $x\notin fv(e_i^{p_i})$ for all $1\leq i\leq|\tilde{\pi}|$
\end{itemize}
By our induction hypothesis we can get:
\begin{itemize}
	\item $\Gamma,\Pi\vdash e^{p''}:(\delta,\kappa)$,
	\item $\Gamma[x^p:(\delta,\kappa)],\Pi\vdash e_i^{p_i}:T_i$ if $s_i=x$
	\item $\Gamma,\Pi\vdash e_i^{p_i}:T_i$ if $s_i\neq x$
\end{itemize}
Where we can then conclude that:
$$\Gamma,\Pi\vdash [\mbox{case}\;e^{p''}\;\tilde{\pi}\;\tilde{o}]^{p}:T$$

		\item[\runa{T-Ref}] Here $e^p=[\mbox{ref}\;e_1^{p_1}]^p$, where
\begin{figure}[H]
	\setlength\tabcolsep{8pt}
	\begin{tabular}{l}
	\runa{T-Ref}\\[0.2cm]
	\inference[]
	{\Gamma,\Pi\vdash  e^{p}:(\delta',\kappa')}
	{\Gamma[\nu x^{p'}:(\delta',\kappa')],\Pi\vdash [\mbox{ref}\;e^{p}]^{p'}:(\emptyset,\kappa)}\\
	\end{tabular}
\end{figure}
Where $\nu x$ is fresh, $\kappa=\{\nu x\}$, and $\Gamma'=\Gamma[x^{p'}:T_1]$.

From our assumption, we know that $x\notin fv(e^p)$.
By \cref{def:fv} we know:
\begin{itemize}
	\item $x\notin fv(e_1^{p_1})$,
\end{itemize}
By our induction hypothesis we can get:
\begin{itemize}
	\item $\Gamma,\Pi\vdash e_1^{p_1}:(\delta',\kappa')$,
\end{itemize}
Where we can then conclude that:
$$\Gamma,\Pi\vdash [\mbox{ref}\;e_1^{p_1}]^{p}:(\emptyset,\kappa)$$

		\item[\runa{T-Ref-read}] Here $e^p=[!e_1^{p_1}]^p$, where
\begin{figure}[H]
	\setlength\tabcolsep{8pt}
	\begin{tabular}{l}
	\runa{T-Ref-read}\\[0.2cm]
	\inference[]
	{\Gamma',\Pi\vdash  e^{p}:(\delta,\kappa)}
	{\Gamma',\Pi\vdash [!e^{p}]^{p'}:T\sqcup(\delta\cup\delta',\emptyset)}\\
	\end{tabular}
\end{figure}
Where $\Gamma'=\Gamma[x^{p'}:T_1]$, $\kappa\neq\emptyset$, $\delta'=\{\nu x^{p'}\mid\nu x\in\kappa\}$, $\nu x_1,\cdots,\nu x_n\in\kappa$.\\ 
	$\{\nu x_1^{p_1},\cdots,\nu x_1^{p_m}\}=uf_{\Upsilon_{p'}}(\nu x_1,\Gamma),\cdots,\{\nu x_n^{p_1'},\cdots,\nu x_n^{p_s'}\}=uf_{\Upsilon_{p'}}(\nu x_n,\Gamma)$, and\\
	$T=\Gamma(\nu x_1^{p_1})\cup\cdots\cup\Gamma(\nu x_1^{p_m})\cup\cdots\cup\Gamma(\nu x_n^{p_1'})\cup\cdots\cup\Gamma(\nu x_n^{p_s'})$.

From our assumption, we know that $x\notin fv(e^p)$.
By \cref{def:fv} we know:
\begin{itemize}
	\item $x\notin fv(e_1^{p_1})$,
\end{itemize}
By our induction hypothesis we can get:
\begin{itemize}
	\item $\Gamma,\Pi\vdash e_1^{p_1}:(\delta',\kappa')$,
\end{itemize}
Where we can then conclude that:
$$\Gamma,\Pi\vdash [!e_1^{p_1}]^{p}:(\delta\cup\delta',\emptyset)$$

		\item[\runa{T-Ref-write}] Here $e^p=[!e_1^{p_1}]^p$, where
\begin{figure}[H]
	\setlength\tabcolsep{8pt}
	\begin{tabular}{l}
	\runa{T-Ref-write}\\[0.2cm]
	\inference[]
	{
		\Gamma',\Pi\vdash  e_1^{p_1}:(\delta,\kappa)&\\
		\Gamma',\Pi\vdash  e_2^{p_2}:(\delta_2,\kappa_2)
	}
	{\Gamma'',\Pi\vdash [e_1^{p_1}\;:=\;e_2^{p_2}]^{p'}:(\delta,\emptyset)}\\
	\end{tabular}
\end{figure}
Where $\Gamma'=\Gamma[x^{p'}:T_1]$, $\Gamma''=\Gamma'[\nu x_1:(\delta_2,\kappa_2),\cdots,\nu x_n:(\delta_2,\kappa_2)]$ and $\nu x_1,\cdots,\nu x_n\in\{\nu x\mid\nu x\in\kappa\}$

From our assumption, we know that $x\notin fv(e^p)$.
By \cref{def:fv} we know:
\begin{itemize}
	\item $x\notin fv(e_1^{p_1})$,
	\item $x\notin fv(e_2^{p_2})$,
\end{itemize}
By our induction hypothesis we can get:
\begin{itemize}
	\item $\Gamma,\Pi\vdash e_1^{p_1}:(\delta,\kappa)$,
	\item $\Gamma,\Pi\vdash e_1^{p_1}:(\delta_2,\kappa_2)$,
\end{itemize}
Where we can then conclude that:
$$\Gamma_1,\Pi\vdash [e_1^{p_1}\;:=\;e_2^{p_2}]^{p}:(\delta,\emptyset)$$
Where $\Gamma_1=\Gamma[\nu x_1:(\delta_2,\kappa_2),\cdots,\nu x_n:(\delta_2,\kappa_2)]$

	\end{description}
\end{proof}

\subsection{Soundness}\label{app:SoundnessProof}
Here, we present the proof for the soundness of the type system.

\begin{theorem}[Soundness of type system]
	Suppose $e^{p'}$ is an occurrence where
	\begin{itemize}
		\item $env\vdash\left\langle e^{p'},sto,(w,\sqsubseteq_w),p\right\rangle\rightarrow\left\langle v,sto',(w',\sqsubseteq_w'),(L,V),p''\right\rangle$,
		\item $\Gamma,\Pi\vdash e^{p'} : T$
		\item $\Gamma,\Pi\vdash env$
		\item $(env,sto,(w,\sqsubseteq_w))\models(\Gamma,\Pi)$
	\end{itemize}
	Then we have that:
	\begin{itemize}
		\item $\Gamma,\Pi\vdash v:T$
		\item $(env,sto',(w',\sqsubseteq_w'))\models(\Gamma,\Pi)$
		\item $(env,(w',\sqsubseteq_w'),v,(L,V))\models(\Gamma,T)$
	\end{itemize}
\end{theorem}

\begin{proof}
	The proof proceeds by induction on the height of the derivation tree for 
	$$env\vdash\left\langle e^{p'},sto,\psi,p\right\rangle\rightarrow\left\langle v,sto',\psi',(L,V),p''\right\rangle$$
	In the base case we have the \runa{Const}, \runa{Var}, \runa{Abs} rules:
	\begin{description}
		\item[\runa{Const}] Here $e^{p'}=c^{p'}$, where
\begin{figure}[H]
	\setlength\tabcolsep{8pt}
	\begin{tabular}{l}
		\runa{Const}\\[0.2cm]
			\inference[]{}
			{env\vdash \left\langle c^{p'},sto,(w,\sqsubseteq_w),p \right\rangle \rightarrow \left\langle c,sto,(w,\sqsubseteq_w),(\emptyset,\emptyset),p' \right\rangle}
	\end{tabular}
\end{figure}
And from our assumptions, we have:
\begin{itemize}
	\item $\Gamma,\Pi\vdash c^{p'} : T$
	\item $\Gamma,\Pi\vdash env$
	\item $(env,sto,(w,\sqsubseteq_w))\models(\Gamma,\Pi)$
\end{itemize}
To type the occurrence $c^{p'}$ we use the rule \runa{T-Const}:
\begin{figure}[H]
	\setlength\tabcolsep{8pt}
	\begin{tabular}{l}
		\runa{T-Const}\\[0.2cm]
			\inference[]{}
			{\Gamma,\Pi\vdash c^{p'}:(\emptyset,\emptyset)}
	\end{tabular}
\end{figure}
We need to show that \cat{1)} $\Gamma,\Pi\vdash c:T$, \cat{2)} $(env,sto',(w',\sqsubseteq_w'))\models(\Gamma,\Pi)$, and \cat{3)} $(env,v,(w',\sqsubseteq_w'),(L,V))\models(\Gamma,T)$.
\begin{description}
	\item[1)] Since we know that the value is a constant, we need to use the \runa{Constant} rule:
		\begin{figure}[H]
			\setlength\tabcolsep{8pt}
			\begin{tabular}{l}
				\runa{Constant}\\[0.2cm]
					\inference[]{}
					{\Gamma,\Pi\vdash c:(\delta,\emptyset)}
			\end{tabular}
		\end{figure}
		Since both the dependency pair and type for constants are $(\emptyset,\emptyset)$, we can then conclude that $\delta=\emptyset$.

	\item[2)] Since there are no updates to $sto$, $(w,\sqsubseteq_w)$, we then know from our assumptions that $(env,sto,(w,\sqsubseteq_w))\models(\Gamma,\Pi)$ holds after an evaluation.

	\item[3)] Since there are no updates to $sto$, $(w,\sqsubseteq_w)$, and that the dependency pair and type are $(\emptyset,\emptyset)$, we then know from \cref{def:TAgree} that $(env,v,(w',\sqsubseteq_w'),(\emptyset,\emptyset))\models(\Gamma,(\emptyset,\emptyset))$ holds.
\end{description}

		\item[\runa{Abs}] Here $e^{p'}=[\lambda\;x.e^{p''}]^{p'}$, where
\begin{figure}[H]
	\setlength\tabcolsep{8pt}
	\begin{tabular}{l}
		\input{sections/appendix/ColRules/abs.tex}
	\end{tabular}
\end{figure}
And from our assumptions, we have:
\begin{itemize}
	\item $\Gamma,\Pi\vdash [\lambda\;x.e^{p''}]^{p'} : T$
	\item $\Gamma,\Pi\vdash env$
	\item $(env,sto,(w,\sqsubseteq_w))\models(\Gamma,\Pi)$
\end{itemize}
To type the occurrence $\left[\lambda\;x.e^{p''}\right]^{p'}$ we use the rule \runa{T-Abs}:
\begin{figure}[H]
	\setlength\tabcolsep{8pt}
	\begin{tabular}{l}
		\runa{T-Abs}\\[0.2cm]
			\inference[]
				{\Gamma[x^{p_0}:T_1],\Pi\vdash  e_1^{p_1}:T_2}
				{\Gamma,\Pi\vdash  \left[\lambda\;x.e_1^{p_1}\right]^{p'}:T_1\rightarrow T_2}
	\end{tabular}
\end{figure}
Where $p'\sqsubseteq_\Pi p_0\wedge p_0\sqsubseteq_\Pi p$.

We need to show that \cat{1)} $\Gamma,\Pi\vdash [\lambda\;x.e^{p''}]:T$, \cat{2)} $(env,sto',(w',\sqsubseteq_w'))\models(\Gamma,\Pi)$, and \cat{3)} $(env,v,(w',\sqsubseteq_w'),(L,V))\models(\Gamma,T)$.
\begin{description}
	\item[1)] Since the value \runa{Abs} evaluates to is a closure, we must use the \runa{Closure} rule:
		\begin{figure}[H]
			\setlength\tabcolsep{8pt}
			\begin{tabular}{l}
				\runa{Closure}\\[0.4cm]
					\inference[]
					{
						\Gamma,\Pi\vdash env \\
						\Gamma[x^{p_0}:T_1],\Pi\vdash e_1^{p_1}:T_2
					}
					{\Gamma,\Pi\vdash \left\langle x^{p_0}, e_1^{p_1}, env \right\rangle:T_1\rightarrow T_2}
			\end{tabular}
		\end{figure}
		We get the first premise from our assumption, and the second premise we can get from the first premise from \runa{T-Abs}.
		The closure type we get from \runa{T-Abs}.

	\item[2)] Since there are no updates to $sto$ and $(w,\sqsubseteq_w)$, we then know from our assumptions that $(env,sto,(w,\sqsubseteq_w))\models(\Gamma,\Pi)$ holds after an evaluation.

	\item[3)] Since there are no updates to $sto$, $(w,\sqsubseteq_w)$, and that the dependency pair is $(\emptyset,\emptyset)$, we then know from \cref{def:TAgree} that $(env,v,(w',\sqsubseteq_w'),(\emptyset,\emptyset))\models(\Gamma,(\emptyset,\emptyset))$ holds.
\end{description}

	\end{description}

	Next, follows the induction step:
	\begin{description}
		\item[\runa{App}] Here $e^{p'}=\left[e_1^{p'}\;e_2^{p''}\right]^{p'}$, where
\begin{figure}[H]
	\setlength\tabcolsep{8pt}
	\begin{tabular}{l}
		\input{sections/appendix/ColRules/app.tex}
	\end{tabular}
\end{figure}

And from our assumptions, we have that:
\begin{itemize}
	\item $\Gamma,\Pi\vdash \left[e_1^{p_1}\;e_2^{p_2}\right]^{p'}:T$,
	\item $\Gamma,\Pi\vdash env$
	\item $(env,sto,(w,\sqsubseteq_w))\models(\Gamma,\Pi)$,
\end{itemize}
To type $[e_1^{p_1}\;e_2^{p_2}]^{p'}$ we need to use the \runa{T-App} type rule, where we have:
\begin{figure}[H]
	\setlength\tabcolsep{8pt}
	\begin{tabular}{l}
		\runa{T-App}\\[0.2cm]
			\inference[]
			{\Gamma,\Pi\vdash e_1^{p_1}:T_1\rightarrow T &\\
			\Gamma,\Pi\vdash e_2^{p_2}:T_1}
			{\Gamma,\Pi\vdash [e_1^{p_1} \; e_2^{p_2}]^{p'}:T}
	\end{tabular}
\end{figure}
We must show that \cat{1)} $\Gamma,\Pi\vdash v:T$, \cat{2)} $(env,sto',(w',\sqsubseteq_w'))\models(\Gamma,\Pi)$, and \\
\cat{3)} $(env,v,(w',\sqsubseteq_w'),(L,V))\models(\Gamma,T)$.

\begin{description}
	\item[1)] To conclude, we first need to look at the premises.
	Due to the first premise: since from our assumptions we have $(env,sto,(w,\sqsubseteq_w))\models(\Gamma,\Pi)$ and $\Gamma,\Pi\models env$, and from the first premise of \runa{T-App}, we can get the following our induction hypothesis:
	\begin{itemize}
		\item $\Gamma,\Pi\vdash v_1:T_1\rightarrow T$,
		\item $(env,sto_1,(w_1,\sqsubseteq_w^1))\models(\Gamma,\Pi)$,
		\item $(env,v,(w_1,\sqsubseteq_w^1),(L_1,V_1))\models(\Gamma,T_1\rightarrow T)$
	\end{itemize}

	Due to the first premise and the second premise of \runa{T-App} we can get the following our induction hypothesis:
	\begin{itemize}
		\item $\Gamma,\Pi\vdash v_2:T_1$,
		\item $(env,sto_2,(w_2,\sqsubseteq_w^2))\models(\Gamma,\Pi)$,
		\item $(env,v,(w_2,\sqsubseteq_w^2),(L_2,V_2))\models(\Gamma,T_1)$
	\end{itemize}

	For the third premise, we first need to look at the value of the first premise, where since we know it is an abstraction, it is concluded by the \runa{Closure} rules:
	\begin{figure}[H]
		\setlength\tabcolsep{8pt}
		\begin{tabular}{l}
			\runa{Closure}\\[0.4cm]
				\inference[]
				{
					\Gamma,\Pi\vdash env' \\
					\Gamma_1,\Pi\vdash e_3^{p_3}:T
				}
				{\Gamma;\Pi\vdash \left\langle x^{p_0}, e_3^{p_3}, env' \right\rangle:T_1\rightarrow T}
		\end{tabular}
	\end{figure}
	Where $\Gamma_1=\Gamma[x^{p_0}:T_1]$.
	We then need to show that there exists a $\Gamma_1$ such that 
	\cat{a)} $\Gamma_1;\Pi\vdash e_3^{p_3}:T$, and 
	\cat{b)} $(env'[x\mapsto v_2],sto_2,(w_3,\sqsubseteq_w^3))\models(\Gamma_1,\Pi)$.
	\begin{description}
		\item[a)] From our assumption, we know that $\Gamma,\Pi\vdash env$ and since we do not allow binding a closure to a location, we then also know that $env=env'[env'']$ for some $env''$, i.e., $env$ contains more bindings than $env'$.
			Based on this, we then also know that $\Gamma,\Pi\vdash env'$, which then conclude the first premise of the \runa{Closure} rule.
	
			For the second premise, we need to show that there exists a $x^{p_0}:T_1$, where from the \runa{T-Abs} rule we know that $p_0\sqsubseteq_\Pi p_3$.
			Since we know that $env[x\mapsto v_2]$, and from the second premise of \runa{T-App} that $\Gamma;\Pi\vdash v_2:T_1$ and $w_2'=w_2[x^{p_2}\mapsto(L_2,V_2)]$, there must be a $x^{p_2}$ such that $p_0=p_2$ and $\Gamma_1=\Gamma[x^{p_2}:T_1]$.

		\item[b)] We know that $(env,sto_2,(w_2,\sqsubseteq_w^2))\models(\Gamma,\Pi)$ from the second premise.
			From \cat{a)} we know that $env=env'[env'']$ as such we also know that $(env',sto_2,(w_2,\sqsubseteq_w^2))\models(\Gamma,\Pi)$.
			We then need to show that $(env'[x\mapsto v_2],sto_2,(w_3,\sqsubseteq_w^2))\models(\Gamma_1,\Pi)$, where $w_3=w_2[x^{p_2}\mapsto(L_2,V_2)]$ and $\Gamma_1=\Gamma[x^{p_2}:T_1]$.
		
			We then only need to show that the updates for $x^{p_2}$ holds.
			Since $x^{p_2}$ is bound in $w_3$, $\Gamma_1$, and $x$ is bound in $env$, that $(env,v,(w_2,\sqsubseteq_w^2),(L_2,V_2))\models(\Gamma,T_1)$, and that $p_2 \sqsubseteq_w^2 p_2$.
			By \cref{def:EnvAgree} we then know that $(env'[x\mapsto v_2],sto_2,(w_3,\sqsubseteq_w^2))\models(\Gamma_1,\Pi)$.
	\end{description}
	From \cat{a)} and \cat{b)}, we can then conclude the third premise of \runa{App} by our induction hypothesis:
	\begin{itemize}
		\item $\Gamma_1,\Pi\vdash v:T$,
		\item $(env'[x\mapsto v_2],sto',(w',\sqsubseteq_w'))\models(\Gamma_1,\Pi)$,
		\item $(env'[x\mapsto v_2],(w',\sqsubseteq_w'),(L_3,V_3))\models(\Gamma,T)$
	\end{itemize}
	As such, since $x^{p_2}\notin fv(v)$ then by \cref{lemma:Strength} we get
	$\Gamma,\Pi\vdash v:T$

\item[2)] Due to \cat{1)} we know that $(env'[x\mapsto v_2],sto',(w',\sqsubseteq_w'))\models(\Gamma_1,\Pi)$, and from \cref{lemma:Strength} we know $x^{p_2}\notin fv(v)$.
	From \cref{lemma:His} we can get the following:
	\begin{itemize}
		\item if $x^{p''}\in dom(w_1)\backslash dom(w)$ then $x^{p''}\notin fv(e_1^{p_1})$
		\item if $x^{p''}\in dom(w_2)\backslash dom(w_1)$ then $x^{p''}\notin fv(e_2^{p_2})$
		\item if $x^{p''}\in dom(w')\backslash dom(w_3)$ then $x^{p''}\notin fv(e_3^{p_3})$
	\end{itemize}
	We then also know that: if $x^{p''}\in dom(w')\backslash dom(w)$ then $x^{p''}\notin fv(e^{p'})$.
	We can then get $(env',sto',(w',\sqsubseteq_w'))\models(\Gamma,\Pi)$.
	Since $env=env'[env'']$ and we know that $(env,sto',(w_3,\sqsubseteq_w^2))\models(\Gamma,\Pi)$, due to \cref{def:EnvAgree} since $w_3=w_2[x^{p_2}\mapsto(L_2,V_2)]$ and $x^{p_2}\notin dom(\Gamma)$.
	We can then get:
	$$(env,sto',(w',\sqsubseteq_w'))\models(\Gamma,\Pi)$$

\item[3)] Due to \cat{1)} and \cat{2)} we can then immediatly conclude that:
	$$(env,v,(w',\sqsubseteq_w'),(L,V))\models(\Gamma,T)$$
\end{description}

		\item[\runa{App-const}] Here $e^{p'}=[ c\;e_1^{p'}\;e_2^{p''}]^{p'}$, where
\begin{figure}[H]
	\setlength\tabcolsep{8pt}
	\begin{tabular}{l}
		\input{sections/appendix/ColRules/appconst.tex}
	\end{tabular}
\end{figure}

And from our assumptions, we have that:
\begin{itemize}
	\item $\Gamma,\Pi\vdash \left[e_1^{p_1}\;e_2^{p_2}\right]^{p'}:T$,
	\item $\Gamma,\Pi\vdash env$
	\item $(env,sto,(w,\sqsubseteq_w))\models(\Gamma,\Pi)$,
\end{itemize}

To type $[c\;e_1^{p_1}\;e_2^{p_2}]^{p'}$ we need to use the \runa{T-App-const} rule, where we have:
\begin{figure}[H]
	\setlength\tabcolsep{8pt}
	\begin{tabular}{l}
	\inference[]
	{
		\Gamma,\Pi\vdash e_1^{p}:(\delta_1,\emptyset) &\\
		\Gamma,\Pi\vdash e_2^{p'}:(\delta_2,\emptyset)
	}
	{\Gamma,\Pi\vdash [c\;e_1^{p} \; e_2^{p'}]^{p''}:(\delta_1\cup\delta_2,\emptyset)}
	\end{tabular}
\end{figure}
Where $c$ is a functional constant.
We must show that \cat{1)} $\Gamma,\Pi\vdash v:T$, \\
\cat{2)} $(env,sto',(w',\sqsubseteq_w'))\models(\Gamma,\Pi)$, and \cat{3)} $(env,v,(w',\sqsubseteq_w'),(L,V))\models(\Gamma,T)$.

\begin{description}
	\item[1)] As we know that the application of all functional constants are constants, we know that the type of the value must be concluded by \runa{Constant}.
	\begin{figure}[H]
		\setlength\tabcolsep{8pt}
		\begin{tabular}{l}
		\runa{Constant}\\[0.2cm]
			\inference[]{}
				{\Gamma,\Pi\vdash  c:(\delta, \emptyset)}\\[1cm]
		\end{tabular}
	\end{figure}
	Where we know that the type can contain occurrences used.
	From the type rule \runa{T-App-const} we first need to look at the premises.
	Due to the first premise: since from our assumptions we have $(env,sto,(w,\sqsubseteq_w))\models(\Gamma,\Pi)$ and $\Gamma,\Pi\models env$, and from the first premise of \runa{T-App-const}, we can get the following our induction hypothesis:
	\begin{itemize}
		\item $\Gamma,\Pi\vdash v_1:T_1\rightarrow T$,
		\item $(env,sto_1,(w_1,\sqsubseteq_w^1))\models(\Gamma,\Pi)$,
		\item $(env,v,(w_1,\sqsubseteq_w^1),(L_1,V_1))\models(\Gamma,T_1\rightarrow T)$
	\end{itemize}

	Due to the first premise and the second premise of \runa{T-App-const} we can get the following our induction hypothesis:
	\begin{itemize}
		\item $\Gamma,\Pi\vdash v_2:T_1$,
		\item $(env,sto_2,(w_2,\sqsubseteq_w^2))\models(\Gamma,\Pi)$,
		\item $(env,v,(w_2,\sqsubseteq_w^2),(L_2,V_2))\models(\Gamma,T_1)$
	\end{itemize}
	As we can see from the rule \runa{T-App-const}, we take an union of $\delta_1$ and $\delta_2$, we can conclude that: $\Gamma,\Pi\vdash v:T$.

\item[2)] Due to \cat{1)}, we can conclude that: $(env,sto',(w',\sqsubseteq_w'))\models(\Gamma,\Pi)$.

\item[3)] Due to \cat{1)}, and \cat{2)}, and that the dependency pair is an union of the dependencies from the premises, and the type is an union of the set of occurrences in the premises, 
	we then know from \cref{def:TAgree} that: $(env,v,(w',\sqsubseteq_w'),(L,V))\models(\Gamma,T)$.
\end{description}

		\item[\runa{App-rec}] Here $e^{p'}=\left[e_1^{p'}\;e_2^{p''}\right]^{p'}$, where
\begin{figure}[H]
	\setlength\tabcolsep{8pt}
	\begin{tabular}{l}
		\input{sections/appendix/ColRules/apprec.tex}
	\end{tabular}
\end{figure}

And from our assumptions, we have that:
\begin{itemize}
	\item $\Gamma,\Pi\vdash \left[e_1^{p_1}\;e_2^{p_2}\right]^{p'}:T$,
	\item $\Gamma,\Pi\vdash env$
	\item $(env,sto,(w,\sqsubseteq_w))\models(\Gamma,\Pi)$,
\end{itemize}
To type $[e_1^{p_1}\;e_2^{p_2}]^{p'}$ we need to use the \runa{T-App} type rule, where we have:
\begin{figure}[H]
	\setlength\tabcolsep{8pt}
	\begin{tabular}{l}
		\runa{T-App}\\[0.2cm]
			\inference[]
			{\Gamma,\Pi\vdash e_1^{p_1}:T_1\rightarrow T &\\
			\Gamma,\Pi\vdash e_2^{p_2}:T_1}
			{\Gamma,\Pi\vdash [e_1^{p_1} \; e_2^{p_2}]^{p'}:T}
	\end{tabular}
\end{figure}
The rest of the proof for this case follows \runa{App}.

		\item[\runa{Let}] Here $e^{p'}=\left[\mbox{let}\;x\;e_1^{p_1}\;e_2^{p_2}\right]^{p'}$, where
\begin{figure}[H]
	\setlength\tabcolsep{8pt}
	\begin{tabular}{l}
		\input{sections/appendix/ColRules/let.tex}
	\end{tabular}
\end{figure}
From our assumption, we know that 
\begin{itemize}
	\item $\Gamma,\Pi\vdash \left[\mbox{let}\;x\;e_1^{p_1}\;e_2^{p_2}\right]^{p'}:T$,
	\item $\Gamma,\Pi\vdash env$
	\item $(env,sto,(w,\sqsubseteq_w))\models(\Gamma,\Pi)$,
\end{itemize}
To type $[\mbox{let}\;x\;e_1^{p_1}\;e_2^{p_2}]^{p'}$ we have two choices, depending on what $e_1^{p_1}$ evaluates to, i.e., if $v_1=\loc$ or $v_1\neq\loc$.
\begin{description}
	\item[I)] If $e_1^{p_1}$ evaluates to a location, $v_1=\loc$, we need to use the \runa{T-Let-1} rule, since the binding is an alias to $\loc$, where we have:
		\begin{figure}[H]
			\setlength\tabcolsep{8pt}
			\begin{tabular}{l}
				\runa{T-Let-1}\\[0.2cm]
					\inference[]
					{\Gamma,\Pi\vdash e_1^{p_1}:(\delta,\kappa) &\\
					\Gamma_1,\Pi\vdash e_2^{p_2}:T}
					{\Gamma,\Pi\vdash [\mbox{let}\; x \; e_1^{p_1} \; e_2^{p_2}]^{p'}:T}
			\end{tabular}
		\end{figure}
		Where $\kappa\neq\emptyset$, and $\Gamma_1=\Gamma[x^{p_1}:(\delta,\kappa\cup\{x\})]$.
		From our assumptions we have $(env,sto,(w,\sqsubseteq_w))\models(\Gamma,\Pi)$ and $\Gamma,\Pi\models env$, and due to the first premises of \runa{Let} and \runa{T-Let-1}, we can then get from our induction hypothesis:
	\begin{itemize}
		\item $\Gamma,\Pi\vdash v_1:(\delta,\kappa)$, where $\kappa\neq\emptyset$,
		\item $(env,sto_1,(w_1,\sqsubseteq_w^1))\models(\Gamma,\Pi)$,
		\item $(env,v,(w_1,\sqsubseteq_w^1),(L_1,V_1))\models(\Gamma,(\delta,\kappa))$
	\end{itemize}
	Since the first premise evaluates to a location, we then also know that $\kappa\neq\emptyset$, and as such, the value is concluded by \runa{Location}.

	In the second premise, we first need to show that \cat{a)} $\Gamma[x^{p_1}:(\delta,\kappa\cup\{x\})],\Pi\vdash env[x\mapsto v_1]$ and \cat{b)} $(env[x\mapsto v_1],sto_1,(w_2,\sqsubseteq_w^1))\models(\Gamma_1,\Pi)$.
	\begin{description}
		\item[a)] Since we know from the first premise that $\Gamma,\Pi\vdash v_1:(\delta,\kappa)$, $\Gamma[x^{p_1}:(\delta,\kappa\cup\{x\})]$, and $env[x\mapsto v_1]$, we then also know from \cref{def:TEnv} that 
			$$\Gamma_1,\Pi\vdash env[x\mapsto v_1]$$
		\item[b)] Due to \runa{a)} , $(env,sto_1,(w_1,\sqsubseteq_w^1))\models(\Gamma,\Pi)$, and \cref{def:TAgree} we only need to show that $(env[x\mapsto v_1],(w_2,\sqsubseteq_w^1),\loc)\models(\Gamma_1,\kappa\cup\{x\})$.

		Since, due to the first premise, we know that $(env,v,(w_2,\sqsubseteq_w^1),\loc)\models(\Gamma,\kappa)$ must hold.
		We also know that if $\kappa^0$ is a good partition, then there must exists a $\kappa_i^0\in\kappa^0$, such that $x\in\kappa_i^0$ and there must also be a $\nu x\in\kappa$ such that $\nu x\in\kappa_i^0$.
		Based on this we know that $(env[x\mapsto v_1],(w_2,\sqsubseteq_w^1),\loc)\models(\Gamma[x^{p_1}:(\delta,\kappa\cup\{x\})],\kappa)$ must also hold.
	\end{description}
	From this, we can then use our induction hypothesis on the second premise, where we then get:
	\begin{itemize}
		\item $\Gamma_1,\Pi\vdash v:T$,
		\item $(env[x\mapsto v_1],sto',(w',\sqsubseteq_w'))\models(\Gamma_1,\Pi)$,
		\item $(env[x\mapsto v_1],(w',\sqsubseteq_w'),(L,V))\models(\Gamma_1,T)$
	\end{itemize}
	From \cref{lemma:Strength}, we know that $x^{p_1}\notin fv(v)$ and by \cref{lemma:His} we know that: if $x^{p''}\in dom(w')\backslash dom(w)$ then $x^{p''}\notin fv(e^{p'})$.
	We can the conclude, for this case:
	\begin{itemize}
		\item $\Gamma,\Pi\vdash v:T$,
		\item $(env,sto',(w',\sqsubseteq_w'))\models(\Gamma,\Pi)$,
		\item $(env,v,(w',\sqsubseteq_w'),(L,V))\models(\Gamma,T)$
	\end{itemize}

	\item[II)]	If $e_1^{p_1}$ does not evaluate to a location, we need to use the \runa{T-Let-2} type rule, where we have:
		\begin{figure}[H]
			\setlength\tabcolsep{8pt}
			\begin{tabular}{l}
				\runa{T-Let-2}\\[0.2cm]
					\inference[]
					{\Gamma,\Pi\vdash e_1^{p_1}:T_1 &\\
					\Gamma_1,\Pi\vdash e_2^{p_2}:T}
					{\Gamma,\Pi\vdash [\mbox{let}\; x \; e_1^{p_1} \; e_2^{p_2}]^{p'}:T}
				\end{tabular}
			\end{figure}
			Where $\Gamma_1=\Gamma[x^{p_1}:T_1]$.
			From our assumptions we have $(env,sto,(w,\sqsubseteq_w))\models(\Gamma,\Pi)$ and $\Gamma,\Pi\models env$, and due to the first premises of \runa{Let} and \runa{T-Let-1}, from our induction hypothesis we can get:
		\begin{itemize}
			\item $\Gamma,\Pi\vdash v_1:T_1$,
			\item $(env,sto_1,(w_1,\sqsubseteq_w^1))\models(\Gamma,\Pi)$,
			\item $(env,v,(w_1,\sqsubseteq_w^1),(L_1,V_1))\models(\Gamma,T_1)$
		\end{itemize}
		We also know that the type of $v_1$ is not a location, because then it would be concluded by \cat{I)}.
		In the second premise, we first need to show that \cat{a)} $\Gamma_1,\Pi\vdash env[x\mapsto v_1]$ and \\
		\cat{b)} $(env[x\mapsto v_1],sto_1,(w_2,\sqsubseteq_w^1))\models(\Gamma_1,\Pi)$.
		\begin{description}
			\item[a)] Since we know from the first premise that $\Gamma,\Pi\vdash v_1:T_1$, $\Gamma_1=\Gamma[x^{p_1}:T_1]$ and $env[x\mapsto v_1]$, due to \cref{def:TEnv} we then know that:
				$$\Gamma_1,\Pi\vdash env[x\mapsto v_1]$$
			\item[b)] Due to \cat{a)}, we know know that since we bind $x$ in $env$, $x^{p_1}$ in $w$ and $\Gamma$, we then know that $(env[x\mapsto v_1],sto_1,(w_2,\sqsubseteq_w^1))\models(\Gamma_1,\Pi)$.
		\end{description}
		From \cat{a)} \cat{b)} we can then use the induction hypothesis
		\begin{itemize}
			\item $\Gamma_1,\Pi\vdash v:T$,
			\item $(env[x\mapsto v_1],sto',(w',\sqsubseteq_w'))\models(\Gamma_1,\Pi)$,
			\item $(env,v,(w',\sqsubseteq_w'),(L,V))\models(\Gamma_1,T)$
		\end{itemize}
		From \cref{lemma:Strength}, we know that $x^{p_1}\notin fv(v)$ and by \cref{lemma:His} we know that: if $x^{p''}\in dom(w')\backslash dom(w)$ then $x^{p''}\notin fv(e^{p'})$.
		We can the conclude, for this case:
		\begin{itemize}
			\item $\Gamma,\Pi\vdash v:T$,
			\item $(env,sto',(w',\sqsubseteq_w'))\models(\Gamma,\Pi)$,
			\item $(env,v,(w',\sqsubseteq_w'),(L,V))\models(\Gamma,T)$
		\end{itemize}
\end{description}

		\item[\runa{Let-rec}] Here $e^{p'}=\left[\mbox{let}\;x\;e_1^{p_1}\;e_2^{p_2}\right]^{p'}$, where
\begin{figure}[H]
	\setlength\tabcolsep{8pt}
	\begin{tabular}{l}
		\input{sections/appendix/ColRules/letrec.tex}
	\end{tabular}
\end{figure}
From our assumption, we know that 
\begin{itemize}
	\item $\Gamma,\Pi\vdash \left[\mbox{let}\;x\;e_1^{p_1}\;e_2^{p_2}\right]^{p'}:T$,
	\item $\Gamma,\Pi\vdash env$
	\item $(env,sto,(w,\sqsubseteq_w))\models(\Gamma,\Pi)$,
\end{itemize}
To type $[\mbox{let}\;x\;e_1^{p_1}\;e_2^{p_2}]^{p'}$ we need to use the \runa{T-Let-2} type rule:
\begin{figure}[H]
	\setlength\tabcolsep{8pt}
	\begin{tabular}{l}
		\runa{T-Let-rec}\\[0.2cm]
			\inference[]
			{\Gamma,\Pi\vdash e_1^{p}:T_1\rightarrow T_2 &\\
			\Gamma[f^p:T_1\rightarrow T_2],\Pi\vdash e_2^{p}:T}
			{\Gamma,\Pi\vdash [\mbox{let rec}\; f \; e_1^{p} \; e_2^{p'}]^{p''}:T}\\[1cm]
	\end{tabular}
\end{figure}
Where going to denote $env'=[f\mapsto\left\langle x,f,e_1^{p_1},env'\rangle\right]$ and $\Gamma'=\Gamma[f^p:T_1\rightarrow T_2]$.
From our assumptions we have $(env,sto,(w,\sqsubseteq_w))\models(\Gamma,\Pi)$ and $\Gamma,\Pi\models env$, and due to the first premises of \runa{Let-rec} and \runa{T-Let-rec}, we can from our induction hypothesis get:
\begin{itemize}
	\item $\Gamma,\Pi\vdash v_1:T_1$,
	\item $(env,sto_1,(w_1,\sqsubseteq_w^1))\models(\Gamma,\Pi)$,
	\item $(env,v,(w_1,\sqsubseteq_w^1),(L_1,V_1))\models(\Gamma,T_1)$
\end{itemize}
In the second premise, we first need to show that \cat{a)} $\Gamma',\Pi\vdash env'$\\
	\cat{b)} $(env',sto_1,(w_2,\sqsubseteq_w^1))\models(\Gamma',\Pi)$.
	\begin{description}
		\item[a)] Since we know from the first premise that $\Gamma,\Pi\vdash v_1:T_1$, and we know that\\
			$\Gamma_1=\Gamma[x^{p_1}:T_1]$ and $env[x\mapsto v_1]$.
			Due to \cref{def:TEnv} we then know that:
			$$\Gamma_1,\Pi\vdash env[x\mapsto v_1]$$
		\item[b)] Since we bind $x$ in $env$, $x^{p_1}$ in $w$ and $\Gamma$, and due to \cat{a)} we then know that $(env[x\mapsto v_1],sto_1,(w_2,\sqsubseteq_w^1))\models(\Gamma_1,\Pi)$.
	\end{description}
	From \cat{a)} \cat{b)} we can then use the induction hypothesis
	\begin{itemize}
		\item $\Gamma_1,\Pi\vdash v:T$,
		\item $(env[x\mapsto v_1],sto',(w',\sqsubseteq_w'))\models(\Gamma_1,\Pi)$,
		\item $(env,,v(w',\sqsubseteq_w'),(L,V))\models(\Gamma_1,T)$
	\end{itemize}
	From \cref{lemma:Strength}, we know that $x^{p_1}\notin fv(v)$ and by \cref{lemma:His} we know that: if $y^{p''}\in dom(w')\backslash dom(w)$ then $y^{p''}\notin fv(e^{p'})$.
	We can the conclude, for this case:
	\begin{itemize}
		\item $\Gamma,\Pi\vdash v:T$,
		\item $(env,sto',(w',\sqsubseteq_w'))\models(\Gamma,\Pi)$,
		\item $(env,v,(w',\sqsubseteq_w'),(L,V))\models(\Gamma,T)$
	\end{itemize}

		\item[\runa{ref}] Here $e^{p'}=\left[\mbox{ref}\;e_1^{p_1}\right]^{p'}$, where
\begin{figure}[H]
	\setlength\tabcolsep{8pt}
	\begin{tabular}{l}
		\input{sections/appendix/ColRules/ref.tex}
	\end{tabular}
\end{figure}
From our assumption, we know that 
\begin{itemize}
	\item $\Gamma,\Pi\vdash \left[\mbox{ref}\;e_1^{p_1}\right]^{p'}:T$,
	\item $\Gamma,\Pi\vdash env$
	\item $(env,sto,(w,\sqsubseteq_w))\models(\Gamma,\Pi)$,
\end{itemize}
To type $[\mbox{ref}\;e_1^{p_1}]^{p'}$ we need to use the \runa{T-Ref} type rule:
\begin{figure}[H]
	\setlength\tabcolsep{8pt}
	\begin{tabular}{l}
		\runa{T-Ref}\\[0.2cm]
			\inference[]
				{\Gamma,\Pi\vdash  e_1^{p_1}:(\delta',\kappa')}
				{\Gamma[\nu x^{p'}:(\delta',\kappa')],\Pi\vdash [\mbox{ref}\;e_1^{p_1}]^{p'}:(\emptyset,\kappa)}
	\end{tabular}
\end{figure}
Where $\kappa=\{\nu x\}$.
We need to show that \cat{1)} $\Gamma,\Pi\vdash [\lambda\;x.e^{p''}]:T$, \cat{2)} $(env,sto',(w',\sqsubseteq_w'))\models(\Gamma,\Pi)$, and \cat{3)} $(env,v,(w',\sqsubseteq_w'),v,(L,V))\models(\Gamma,T)$.

To conclude, we first need to show for the premise, where due to our assumption and from the premise, we can use the induction hypothesis, where we then get:
\begin{itemize}
	\item $\Gamma,\Pi\vdash v_1:(\delta',\kappa')$,
	\item $(env,sto'',(w'',\sqsubseteq_w'))\models(\Gamma,\Pi)$,
	\item $(env,v,(w'',\sqsubseteq_w'),(L,V))\models(\Gamma,(\delta',\kappa'))$
\end{itemize}
We are also going to denote $\Gamma'=\Gamma[\nu x^{p'}:(\delta',\kappa')]$.

\begin{description}
	\item[1)] Since we know that \runa{Ref} evaluates to a location, we know it should be concluded by \runa{Closure}.
	\begin{figure}[H]
		\setlength\tabcolsep{8pt}
		\begin{tabular}{l}
			\runa{Location}\\[0.2cm]
				\inference[]{}
					{\Gamma,\Pi\vdash  \loc:(\delta, \kappa)}
		\end{tabular}
	\end{figure}
	Where $\kappa\neq\emptyset$.
	From the \runa{T-Ref} rule, we know that the type is $(\emptyset,\{\nu x\})$, we can then conclude that $\delta=\emptyset$ and $\kappa=\{\nu x\}$.
	\item[2)] Since we know that $(env,sto'',(w'',\sqsubseteq_w'))\models(\Gamma,\Pi)$, we then need to show for the extension to $sto''$, $w''$, and $\Gamma'$.
		Due to \cat{1)}, $(env,v,(w'',\sqsubseteq_w'),(L,V))\models(\Gamma,(\delta',\kappa'))$, and since we bind $\loc^{p'}$ in $sto''$ and $\nu x^{p'}$ in $\Gamma$, we can then conclude that $(env,sto',(w',\sqsubseteq_w'))\models(\Gamma',\Pi)$
	\item[3)] Due to \cat{1)} and \cat{2)} we can conclude that $(env,v,(w',\sqsubseteq_w'),(L,V))\models(\Gamma,T)$.
\end{description}

		\item[\runa{Ref-write}] Here $e^{p'}=[e_1^{p_1}:=e_2^{p_2}]^{p'}$, where
\begin{figure}[H]
	\setlength\tabcolsep{8pt}
	\begin{tabular}{l}
		\input{sections/appendix/ColRules/refwrite.tex}
	\end{tabular}
\end{figure}

And from our assumptions, we have that:
\begin{itemize}
	\item $\Gamma,\Pi\vdash [e_1^{p_1}:=e_2^{p_2}]^{p'}:T$,
	\item $\Gamma;\Pi\vdash env$
	\item $(env,sto,(w,\sqsubseteq_w))\models(\Gamma,\Pi)$,
\end{itemize}
To type $[e_1^{p_1}:=e_2^{p_2}]^{p'}$ we need to use the \runa{T-Ref-write} rule, where we have:
\begin{figure}[H]
	\setlength\tabcolsep{8pt}
	\begin{tabular}{l}
		\runa{T-Ref-write}\\[0.2cm]
			\inference[]
				{\Gamma,\Pi\vdash  e_1^{p_1}:(\delta,\kappa)&\\
				\Gamma,\Pi\vdash  e_2^{p_2}:(\delta_2,\kappa_2)}
				{\Gamma',\Pi\vdash [e_1^{p_1}\;:=\;e_2^{p_2}]^{p'}:(\delta,\emptyset)}\\
	\end{tabular}
\end{figure}
Where $\Gamma'=\Gamma[\nu x_1:(\delta_2,\kappa_2),\cdots,\nu x_n:(\delta_2,\kappa_2)]$ and $\nu x_1,\cdots,\nu x_n\in\{\nu x\mid\nu x\in\kappa\}$

We must show that \cat{(1)} $\Gamma',\Pi\vdash v:T$, \cat{(2)} $(env,sto',(w',\sqsubseteq_w'))\models(\Gamma',\Pi)$, and \cat{(3)} $(env,v,(w',\sqsubseteq_w'),(L,V))\models(\Gamma',T)$.

To conclude, we first need to show for the premise, where due to our assumption and from the first premise, we can use the induction hypothesis, where we then get:
\begin{itemize}
	\item $\Gamma,\Pi\vdash \loc:(\delta,\kappa)$,
	\item $(env,sto_1,(w_1,\sqsubseteq_w^1))\models(\Gamma,\Pi)$,
	\item $(env,v,(w_1,\sqsubseteq_w^1),(L,V))\models(\Gamma,(\delta,\kappa))$
\end{itemize}
And from the second premise we can get that:
\begin{itemize}
	\item $\Gamma,\Pi\vdash v:(\delta_2,\kappa_2)$,
	\item $(env,sto_2,(w_2,\sqsubseteq_w^2))\models(\Gamma,\Pi)$,
	\item $(env,v,(w_2,\sqsubseteq_w^2),(L,V))\models(\Gamma,(\delta',\kappa'))$
\end{itemize}

Since we know that $(env,sto_1,(w_1,\sqsubseteq_w^1))\models(\Gamma,\Pi)$, $(env,sto_2,(w_2,\sqsubseteq_w^2))\models(\Gamma,\Pi)$ and we bind $\loc$ to $v$.
We also know that from $(env,v,(w_1,\sqsubseteq_w^1),(L,V))\models(\Gamma,(\delta,\kappa))$, that all internal variable in $\kappa$ agrees with the location $\loc$.
We can then conclude that \cat{2)} $(env,sto',(w',\sqsubseteq_w'))\models(\Gamma',\Pi)$.

\begin{description}
	\item[1)] Since we know that the value is unit, $()$, this must be conclude by \runa{Unit}:
		\begin{figure}[H]
			\setlength\tabcolsep{8pt}
			\begin{tabular}{l}
				\runa{Unit}\\[0.2cm]
					\inference[]{}
					{\Gamma',\Pi\vdash  ():(\delta,\emptyset)}\\[0.5cm]
			\end{tabular}
		\end{figure}
	\item[3)] Due to \cat{1)} and \cat{2)} we can then conclude that $(env,v,(w',\sqsubseteq_w'),(L,V))\models(\Gamma',T)$.
\end{description}

	\end{description}
\end{proof}

\end{document}